\countdef\refno=80 \countdef\secno=85 \countdef\equno=90
\countdef\ceistno=91 
{\def\autoeq{ {\global\advance\count90 by1} \eqno(\the\count90) }
\def\autoeql{ {\global\advance\count90 by1} & (\the\count90) }

\def\autoref{ {\global\advance\refno by 1} \kern -5pt [\the\refno]\kern 2pt}
\def\setup{\count90=0 \count80=0 \count91=0 \count85=0}
\def\vline{{\vrule height8pt depth4pt}\; }
\def\Vline{{\vrule height13pt depth8pt}\; }
\def\dddot#1{{\,\hbox{{\raise 8.5pt\hbox{.}
                     \kern -4pt\raise 8.5pt\hbox{.}
                     \kern -4pt\raise 8.5pt\hbox{.}} \kern -12pt $#1$}}}
\def\circum#1{{ \kern -3.5pt $\hat{\hbox{#1}}$ \kern -3.5pt}}
\magnification1200
\setup
\overfullrule=0pt
\voffset=1.5truecm
\centerline{  }
\line{\hfill \hbox{Edinburgh Preprint 97/3}}
\line{\hfill \hbox{hep-th/9702156}}
\line{\hfill \hbox{19th October 1997}}
\line{\hfill \hbox{Second Revision} }
\vskip 1cm
\centerline{\bf Renormalisation Group Flow and Geodesics in the $O(N)$
Model for Large $N$} 
\vskip 0.7cm
\centerline{Brian P. Dolan\footnote*{Currently on leave of absence
from: Department of Mathematical Physics, St. Patrick's College, Maynooth,
Ireland}}
\vskip .3cm
\centerline{\it Department of Physics and Astronomy }
\centerline{\it The University Edinburgh}
\centerline{\it Scotland}
\vskip .3cm
\centerline{e-mail: bdolan@thphys.may.ie}
\bigskip
\centerline{\bf ABSTRACT}
\medskip
\noindent A metric is introduced on the space of parameters (couplings)
describing the large $N$ limit of the $O(N)$ model in Euclidean space.
The geometry associated with this metric is analysed in the particular case of the infinite
volume limit in 3 dimensions and it is shown that the Ricci curvature
diverges at the ultra-violet (Gaussian) fixed point but is finite
and tends to constant negative curvature at the infra-red
(Wilson-Fisher) fixed point. The renormalisation group flow is
examined in terms of geodesics of the metric. The critical line of
cross-over from the Wilson-Fisher fixed point to the Gaussian fixed point
is shown to be a geodesic but all other renormalisation group trajectories,
which are repulsed from the Gaussian fixed point in the ultra-violet,
are not geodesics. The geodesic flow is interpreted in terms of a
maximisation principle for the relative entropy.
\vfill\eject

\noindent
\leftline{\bf{\S 1 Introduction}}
\bigskip
In this paper the idea of a geometry on the space of field theories will
be investigated, in particular within the context of an exactly soluble 
model - the $O(N)$ model in D-dimensions in the limit of large $N$.  
A geometry on the space of theories was used to great effect by 
Zamolodchikov in the case $D=2$,\autoref\newcount\Zam\Zam=\refno,
where a metric was defined on the 
space of couplings of the theory which led to deep insights into the nature 
of renormalisation flow and cross-over between fixed points (the c-theorem).  
This metric was essentially given by the two point correlators of the 
primary fields of the theory, and has no unique generalisation to $D>2$, 
where the concept of a primary field does not play such a central 
r\circum ole. 
However, a related metric can be defined for $D>2$, where the components 
are given by two point correlators of the composite operators associated 
with the couplings \autoref\newcount\DenjoeChris\DenjoeChris=\refno.
The concept goes back much further 
in the statistical mechanics literature and can be traced to ideas of 
Fisher and Rao 
\autoref\newcount\Fisher\Fisher=\refno
\autoref\newcount\Rao\Rao=\refno - the Fisher information  
matrix and \lq\lq relative entropy".  
In the context of ordinary statistics there is a book on the subject\autoref
\newcount\Amari\Amari=\refno
and metrics on the space of thermodynamic states have been investigated in 
some detail by Ruppeiner and Weinhold\autoref
\newcount\Ruppeiner\Ruppeiner=\refno,\autoref.
\newcount\Weinhold\Weinhold=\refno
A related metric in quantum mechanics has been proposed by Provost and 
Vallee,\autoref.
\newcount\PV\PV=\refno
Zamolodchikov
appears to have been the first to use the idea in field theory, albeit only 
for 2 dimensional field theories. Many attempts have been made to  
generalise Zamolodchikov's results to 3 and 4 dimensions\autoref
\newcount\ctheorem\ctheorem=\refno, but these 
have mostly focused on attempts to prove a c-theorem in $D>2$, rather than 
on the intrinsic {\it{geometry}} of the proposed metric.
\bigskip
The geometry itself is also of intrinsic importance. A change
in the way the theory is parameterised would correspond to a general
co-ordinate transformation and non-linear transformations are 
perfectly acceptable, provided all quantities are expressed in a 
manifestly general co-ordinate co-variant manner. For example the
transformation from bare to renormalised parameters, which is in general
non-linear, can be interpreted as a general co-ordinate transformation,
\autoref.
\newcount\GeomRG\GeomRG=\refno
In particular any quantity which is a scalar under general co-ordinate
transformations is automatically independent of the renormalisation
scheme.
\bigskip
The geometry of the space of couplings was taken seriously 
in [\the\DenjoeChris], where Gaussian models (free scalar field 
theories in a finite box) were investigated, and curvatures calculated. 
The concept of a connection on the infinite dimensional space of theories, 
and its r\circum ole in renormalisation group flow, was investigated 
in \autoref\newcount\Sonoda\Sonoda=\refno.  The metric studied in 
[\the\DenjoeChris] was in a sense the 
infra-red limit of the Fourier transform of Zamolodchikov's metric 
(generalised to $D>2$ where the concept of primary fields is not so well 
defined), and so does not contain as much information as that of 
Zamolodchikov, but it is still relevant to an analysis of the long 
distance behaviour of the theory.
\bigskip
The relation between the geometry and the renormalisation flow was 
investigated in \autoref\newcount\geodrg\geodrg=\refno, 
where it was observed that the renormalisation 
flow on the two dimensional space parameterised by the mass and the 
expectation value $\langle \varphi \rangle$ of a scalar field is geodesic 
for free fields, provided $\langle \varphi \rangle = 0$ but not otherwise 
(except in D=2, where all renormalisation trajectories are 
geodesic). 
This was however in the rather restricted cases of free field theories,
where the renormalisation flow is just dictated by canonical dimensions, 
and the slightly less trivial case of the 1-dimensional Ising model.
\bigskip
The purpose of this paper is to pursue these investigations for an 
interacting non-trivial model which is exactly soluble --- the $O(N)$ model 
in the limit of $N \rightarrow \infty$.  For D=3, this model is non-trivial
 and has two fixed points - the Gaussian fixed point (free field theory) 
in the ultra-violet and the non-trivial Wilson-Fisher fixed point in the 
infra-red (which is equivalent to the spherical model
\autoref\newcount\Stanley\Stanley=\refno). 
The Ricci scalar diverges at the Gaussian fixed point
but elsewhere the curvature
is finite, tending to a negative constant in the infra-red.  
It is shown 
that, with the metric used here, the line of cross-over between the Gaussian 
and Wilson-Fisher fixed points is a geodesic and this is related to the 
concept of relative entropy in statistics.
\bigskip
In section 2, the choice of metric that is used will be described,
motivated by considerations of general co-ordinate invariance. 
Section 3 is devoted to the explicit determination of the metric and 
curvature for the $O(N)$ model in D dimensions, for large $N$.  This 
involves the inclusion of $1\over N$ corrections, as the metric proves 
to be degenerate to lowest order.  Section 4 specialises to the infinite 
volume limit in D=3, where it is shown that the Ricci Scalar, 
$R\rightarrow + \infty$ at the Gaussian fixed point, and 
$R\rightarrow - 6\pi^2$ when any of the three parameters of the 
model (constant external source, the mass of the scalar field or 
the 4 - point coupling, $\lambda$) is large.  In particular the infra-red 
fixed point corresponds to $\lambda \rightarrow \infty$.  It is 
also shown that the line of cross-over, from the infra-red to the 
ultra-violet fixed point is a geodesic, and section 5 is devoted to 
an interpretation of this result in terms of relative entropy. Section 6 
contains a summary and conclusions.
\bigskip
There are two appendices, one containing some technical aspects of 
Legendre transforms, which are used in section 3, and a second which 
gives the connection co-efficients, also used in section 3.
\bigskip
\leftline{\bf \S 2 The Metric}
\bigskip
In this section a definition of a metric on the space of couplings will 
be given. The basic motivation follows that of reference 
[\the\DenjoeChris].  Consider a field theory in 
$D$-dimensional Euclidean space
with $n$ couplings $g^a$, $a=1,\ldots,n$, corresponding to 
operators $\hat\Phi_a{(x)}$ 
(in general composite). The definition of the reduced free energy
(i.e. the free energy divided by the temperature) is
$$
W(g) = -\ln Z(g) \quad \hbox{where}\quad Z(g) = \int {\cal D}\varphi 
e^{-S[ \varphi ]}
\autoeq$$
and $S[\varphi]$ is the action. This  gives
$$
1 = \int {\cal D}\varphi e^{-S[ \varphi ]+W} 
\qquad\Rightarrow\qquad dW = \langle dS \rangle,
\autoeq$$
\newcount\norm\norm=\equno
where $dW = \partial_aWdg^a$ is a one-form and $dS = \partial_aSdg^a$ 
can be thought of as an operator valued one-form.  In particular, if the 
action $S$ is linear in the couplings, then 
$$\partial_aS = \int d^Dx \; \hat\Phi_a(x)
\autoeq$$
\newcount\Action\Action=\equno
where $\hat\Phi_a(x)$ is the composite operator associated with the
coupling $g^a$. Thus if $g^{a_0}$ are bare couplings then 
$\hat\Phi_{a_0}(x)$ are bare operators. If one then defines renormalised
couplings $g^{a_R}$, using some preferred scheme, then the renormalised
operators are 
\hbox{$\hat\Phi_{a_R}(x)={(Z^{-1})^{b_0}}_{a_R}\hat\Phi_{b_0}(x)$},
where the operator mixing matrix 
${(Z^{-1})^{b_0}}_{a_R}={\partial g^{b_0} \over \partial g^{a_R}}$ 
is nothing
other than a general co-ordinate transformation matrix for the co-variant
vector with components $\hat\Phi_{a_R}(x)$. Thus the definition
of  $\hat\Phi_a(x)$
given in equation (\the\Action) is a general co-ordinate co-variant
definition even when the action is non-linear in the couplings and is
valid both for bare and renormalised couplings.
Equations (\the\norm) and (\the\Action) are 
referred to as an \lq\lq action Principle" in 
\autoref\newcount\GuidaMagnoli\GuidaMagnoli=\refno.
\bigskip
The metric advocated by O'Connor and Stephens in [\the\DenjoeChris] is 
determined by the infinitesimal line element on the $n$-dimensional
space parameterised by $g^a$ defined by
$$
ds^2 =\langle(dS-dW) \otimes (dS-dW) \rangle.
\autoeq$$
\newcount\linelement\linelement=\equno
In order to be able to pass to the infinite volume limit, it will be 
convenient to divide equation (\the\linelement) by a factor
$V=\int d^Dx$, the volume of space, and use densities. 
Let
$$
\tilde\Phi_a(x) = \hat\Phi_a(x) - \langle\hat\Phi_a(x)\rangle
\autoeq$$
and define
$$
G_{ab} = \int d^Dx \langle\tilde\Phi_a(x)\tilde\Phi_b(0)\rangle.
\autoeq$$
\newcount\metric\metric=\equno
This is the metric which will be investigated here.  Obviously 
$G_{ab} = G_{ba}$ 
and under a general co-ordinate transformation 
$g^a\rightarrow g^{a^\prime}(x)$

$$
\partial_aS\rightarrow\partial_{a^\prime}S= 
{{\partial g^b}\over {\partial g^{a^\prime}}} 
\partial_bS
\autoeq$$
so
$$
G_{ab}\rightarrow G_{a^\prime b^\prime}={{\partial g^c}\over 
{\partial g^{a^\prime}}} 
{{\partial g^d}\over {\partial g^{b^\prime}}} G_{cd}
\autoeq$$
has the correct transformation properties to be considered as a metric.
\bigskip
Of course if bare couplings are used then the $\hat\Phi_a(x)$ are divergent
operators when the regulator is removed. One can either keep the
regulator in place until the end of the calculation or transform
to renormalised operators using a co-ordinate transformation - provided the
formalism is manifestly co-variant it does not matter and
the latter possibility allows a
consistent analysis.
However the R.H.S. of equation (6) contains further 
divergences in general, either infra-red divergences due to the
large $x$-behaviour or ultra-violet divergences due to the small
$x$-behaviour. The usual 
procedure is to perform further subtractions, over and above
any that may have already been used to obtain renormalised operators,
so as to obtain a renormalised 2-point 
function \autoref\newcount\GrahamHugh\GrahamHugh=\refno.  
This will not be done here -- rather $G_{ab}$ 
will be 
defined using a regulator, connections and curvatures will be calculated 
first and only then will the regulator be removed.
There is a good geometrical reason for this strategy. As explained above
multiplicative renormalisation
can be interpreted as a co-ordinate transformation and
so does not change the geometry --- the components of the metric look
different but the geometry (in particular the Ricci scalar) is not
changed. Subtracting extra terms which are non-linear in the
couplings from (\the\metric) would however change
the geometry  and so would change the Ricci scalar. By avoiding such 
subtractions one can be confident that the resulting Ricci scalar
is independent of the renormalisation scheme.
\bigskip
As noted in 
[\the\DenjoeChris], for free field theories, the curvature remains finite
even though the components of the
metric diverge when the regulator is removed.
For the large $N$ limit of the $O(N)$ model in three dimensions, 
it will transpire that the curvature diverges at the Gaussian fixed point
but not elsewhere.
\bigskip
Equation (\the\metric) can be written in a manner more convenient for 
computations. Let
$$
w = {1 \over V} W
\autoeq$$
be the reduced free energy density, so that $W=\int w\;d^Dx$.  Then equation (\the\norm) reads
$$
\partial_aw = {1\over V} \langle\partial_aS\rangle.
\autoeq$$
Differentiating a second time gives
$$
\partial_a\partial_bw={1\over V}
\left\{ \langle \partial_a\partial_bS \rangle - 
\langle\partial_aS\partial_bS\rangle + \langle\partial_aS\rangle 
\langle\partial_bS\rangle \right\}
\autoeq$$
or
$$
G_{ab} = \int d^Dx \langle\tilde\Phi_a(x)\tilde\Phi_b(0)\rangle = 
{1\over V} \langle \partial_a\partial_bS\rangle - \partial_a\partial_bw.
\autoeq$$
\newcount\metricb\metricb=\equno
Despite appearances the right hand side of (\the\metricb) is co-variant under 
general co-ordinate transformations since, if $\partial_bS \rightarrow
\partial_{b^\prime}S= 
{{\partial g^c}\over{\partial g^{b^\prime}}} \partial_cS$ and $\partial_bw 
\rightarrow \partial_{b^\prime}w=
{{\partial g^c}\over{\partial g{b^\prime}}} \partial_cw$ 
then 
$$\partial_{a^\prime}\partial_{b^\prime} S = 
{{\partial g^c}\over{\partial g^{a^\prime}}}  {{\partial g^d}\over
{\partial g^{b^\prime}}}  \partial_c\partial_d S +  {{\partial^2g^c}
\over{\partial g^{a^\prime}\partial g^{b^\prime}}} \partial_c S 
\quad
\hbox{and }
\quad\partial_{a^\prime}\partial_{b^\prime} w = {{\partial g^c}\over
{\partial g^{a^\prime}}}  {{\partial g^d}\over{\partial g^{b^\prime}}}  
\partial_c\partial_d w +  {{\partial^2g^c}\over{\partial g^{a^\prime}
\partial g^{b^\prime}}} \partial_c w. \autoeq$$
So the inhomogeneous terms cancel when expectation values are taken, 
by virtue of equation (\the\norm).  The analysis of this section
has been general co-ordinate co-variant up to this point. Equations
(\the\Action), (\the\metric)
and (\the\metricb) are valid even if renormalised couplings are used
and the action is not linear in the couplings.
If however one chooses parameters in which the action is linear
(these would be the bare parameters of the theory) then equation
(\the\metricb) simplifies to
$$
G_{ab} = - \partial_a\partial_bw.
\autoeq$$
\newcount\Metric\Metric=\equno
The class of such co-ordinate systems is special, of course --- 
only linear co-ordinate transformations are allowed.\footnote*{In some
situations there is a natural complex structure on the space of parameters
and a metric of the form (\the\Metric) can be interpreted as a K\"ahler metric.
The class of allowed co-ordinate transformations which preserve the form
of (\the\Metric) can then be extended to include any complex analytic
transformation. An example is the Seiberg-Witten metric on the parameter 
space of $N=2$ super symmetric Yang-Mills theory in four dimensions
\autoref.}
\newcount\SeibergWitten\SeibergWitten=\refno
Within this class, however, equation (\the\Metric) says that the components 
of the metric can be determined if the partition function, and so $w$, 
is known as a function of the regularised bare parameters. 
For translationally invariant systems, 
this is equivalent to a knowledge of the effective potential, or the 
free energy. Another useful class of co-ordinates is that obtained
by the non-linear co-ordinate transformations associated with the
Legendre transformed variables, these can simplify the metric
even further and will prove useful in the sequel --- this class,
and the resulting form of the metric in terms of the effective potential,
is examined in detail in appendix 1.
\bigskip
When viewed in this light, some singularities in the metric can be given 
a more direct interpretation.  For example, if one of the operators is 
$\varphi^2$ in a scalar field theory, the statistical physics interpretation of the 
co-efficient of ${1\over 2}\varphi^2$, $t = (T-T_c)/T_c$, 
is that it is the deviation from the critical
temperature, and the second 
derivative of the free energy with respect to $t$ is the specific heat, 
hence one expects some components of the metric (\the\Metric) 
to diverge at critical 
points and in general one might expect the curvature to diverge
there also. In fact for the $O(N)$ model at large $N$ in 3 dimensions, 
the critical exponent for the specific heat, 
$\alpha = -1+o\bigl({1\over N}\bigr)$,
is negative at the infra-red (Wilson-Fisher) fixed point, so the specific heat 
is actually finite at $t=0$ and it is only the third derivative of the free 
energy with respect to $t$ that diverges.  Calculation of the Ricci scalar 
however, reveals that it is finite all along the critical line between 
the infra-red and the ultra-violet fixed point, diverging only at the 
ultra-violet (Gaussian) fixed point.  The non-analyticity of the free 
energy along the critical line is still reflected in the Ricci scalar however, 
in that it displays a discontinuity across the critical line.  

\bigskip

\leftline{\bf \S 3 The Geometry of the O(N) Model}
\bigskip
The model that will be investigated here is the O(N) model in D Euclidean 
dimensions, in the limit of $N\rightarrow \infty$.  This is an example of 
a non-trivial interacting field theory (for $D<4$) which can be solved 
exactly.  The model consists of a scalar field $\vec\varphi$ in the vector 
representation of $O(N)$,
 with components $\varphi^i,i=1,\cdots, N$.
The action is (really 
total energy since the space is Euclidean)
$$
S = \int d^D x\biggl\{ {1\over 2}(\bigtriangledown\varphi)^2 + \vec j \cdot 
\vec\varphi + {r\over 2} \varphi^2 + {u\over 4!} (\varphi^2)^2
\biggr\}
\autoeq$$
where $\vec j$ is a constant 
external source and $r$ and $u$ are the bare mass 
and 4-point coupling respectively.
The aim of this section is to determine the metric, as defined in 
the previous section, and to investigate the resulting geometry in terms
of the Levi-Civita connection and the curvature. The variables
$j$, $r$ and $u$ are not particularly convenient for this
purpose and it will prove expedient to transform to an alternative
set, but first we outline the calculation of the partition function
and the effective potential.

The partition function is 
$$
Z[j,r,u] = \int{\cal D} \varphi e^{-S}.
\autoeq$$
\noindent
The form of the scaling function for this model in the limit of large $N$
was investigated in \autoref\newcount\Wallace\Wallace=\refno. The
analysis here will use the steepest 
descents method of 
\autoref\newcount\ZinnJustin\ZinnJustin=\refno, as implemented in 
\autoref\newcount\DenjoeChrisb\DenjoeChrisb=\refno.  
First replace the $\varphi^4$ term with an effective field $\psi$
$$\eqalign{
e^{-\int d^Dx {{u}\over{4!}}(\varphi^2)^2} &= {\cal N}\int {\cal D}\psi 
e^{\int d^Dx\left\{{{N}\over{2}} (\psi - \sqrt{{u}\over{12N}}\varphi^2)^2-u
{{(\varphi^2)^2}\over{4!}}
\right\}}\cr
&= {\cal N}\int {\cal D}\psi e^{\int d^Dx\left\{{{N}\over{2}} \psi^2 - 
\sqrt{{uN}\over{12}}\psi\varphi^2 \right\}}\cr}
\autoeq$$
where ${\cal N}$ is an irrelevant constant, independent of $u$.
Now
$$\eqalign{
Z &= {\cal N}\int{\cal D}\psi {\cal D}\varphi e^{-\int d^Dx\left\{{{1}\over{2}}
(\bigtriangledown\varphi)^2 + \vec j.\vec\varphi + {{1}\over{2}} M^2\varphi^2 - 
{{N}\over{2}} \psi^2\right\}}\cr
&= {\cal N}\int{\cal D}\psi {\cal D}\varphi e^{-\int d^Dx\left\{{{1}\over{2}}
(\bigtriangledown\varphi)^2 + {{1}\over{2}} M^2 \Bigl(\varphi + 
{{j}\over{M^2}}\Bigr)^2-{{j^2}\over{2M^2}}-{{N}\over{2}}\psi^2\right\}} \cr}
\autoeq$$
where $M^2(\psi) = r+\sqrt{{{Nu}\over{3}}}\psi$ is an effective mass for the 
field $\varphi$.  After a shift in $\varphi$, the $\varphi$ integration is 
Gaussian, leading to an effective action for the field $\psi$
$$
Z[j,r,u] = {\cal N}\int{\cal D}\psi e^{-S_{eff}(\psi)}
\autoeq$$
\newcount\Zpsi\Zpsi=\equno
where
$$
S_{eff}(\psi)=-\int d^Dx\biggl\{{{N}\over{2}}\psi^2+{{1}\over{2}}
{{j^2}\over{M^2(\psi)}}\biggr\} + {{N}\over{2}} \int d^Dx \ln 
\{-\bigtriangledown^2+M^2(\psi)\}.
\autoeq$$
\newcount\Seff\Seff=\equno
Note that $Z$ now depends only $j=|\vec j|$, as expected.  So far 
the manipulations are exact, if formal.  The method of steepest 
descents now allows one to evaluate the effective potential as a $1/N$ 
expansion.  Expand $\psi(x)$ around a constant background
$$
\psi(x) = \psi_0 + {{1}\over{\sqrt N}}\varepsilon(x)
\autoeq$$
\newcount\psioh\psioh=\equno
where $\psi_0$ is chosen so that ${{\partial S_{eff}
}\over {\partial \psi(x)}} 
\bigl\vert_{\psi(x)=\psi_0} = 0.$
The function $\int d^Dx \ln \{-\bigtriangledown^2+M^2\}$ will appear so 
frequently in the following, that it will be convenient to give it a name.  
In momentum space
$$
G(m^2):= 
\int{{d^Dp}\over{(2\pi)^D}} \ln(p^2+m^2)\autoeq
$$
\newcount\Gdef\Gdef=\equno
$$
\dot G(m^2):= {{dG}\over{dm^2}} = \int {{d^Dp}\over{(2\pi)^D}} 
{{1}\over{(p^2+m^2)}}\autoeq
$$
\newcount\Gddef\Gddef=\equno
$$
\ddot G(m^2) := {{d^2G}\over{(dm^2)^2}}=- 
\int{{d^Dp}\over{(2\pi)^D}}{{1}\over{(p^2+m^2)^2}}
\autoeq$$
\newcount\Gdddef\Gdddef=\equno
where $m^2(\psi_0)=M^2|_{\psi=\psi_0}=r+\sqrt{Nu\over 3}\psi_0$.
\bigskip
The function $G(m^2)$ is not uniquely specified here, it depends on 
boundary conditions and geometry, for example the integral could 
correspond to infinite Euclidean space or a D-dimensional torus each 
giving different eigenvalues for the Laplacian.  If space is continuous, 
$G(m^2)$ must be rendered finite in some way e.g. by introducing a 
momentum cut-off.  Alternatively, continuous 
space could be replaced by a finite lattice of points and the Laplacian 
becomes a matrix with $\int{{d^Dp}\over{(2\pi)^D}}\rightarrow Tr$.  
In the following we shall work with a generic $G(m^2)$, it being 
understood that different geometries for $D$-dimensional space 
lead to different functions, $G(m^2)$. 

The extremum 
condition now determines $\psi_0$ via
$$
- N\psi_0 + {1 \over 2} {{j^2}\over{m^4}} \sqrt{{Nu}\over{3}} + 
{N\over 2} \dot G(m^2) \sqrt{{Nu}\over{3}} = 0,
\autoeq$$
\newcount\extrem\extrem=\equno
where ${{\partial M^2(\psi(y))}\over{\partial \psi(x)}}  = \sqrt{{Nu}
\over{3}} \delta (x-y)$ has been used.  Equation (\the\extrem) determines 
$\psi_0$ as a function of $j$, $r$ and $u$ since 
$m^2(\psi_0)=r+\sqrt{Nu\over 3}\psi_0$.  
Expanding $S_{eff}$ in equation (\the\Seff) 
around $\psi_0$, using (\the\psioh) gives 
$$
S_{eff} = {N \over 2} V\{G(m^2) - \psi_0^2 - {j^2\over Nm^2}\} + o(\varepsilon^2)
\autoeq$$
where again $V = \int d^Dx$ is the volume of space.
Thus the reduced free energy density
$$
w = -{1 \over V} \ln Z
\autoeq$$
is given (up to an irrelevant constant, independent of $j, r \,\, 
\hbox{and} \,\, u$) by,
$$
w = {N \over 2} \{G(m^2)-{j^2\over Nm^2}
-\psi_0^2\} + o(1)
\autoeq$$
with $m^2$ determined implicitly by equation (\the\extrem).  If we define 
$\lambda = {{Nu}\over{3}}, J = j/\sqrt{N}$ and $\tilde w = {w\over N}$
then
$$
\tilde w  = {1 \over 2} \{G(m^2)-J^2/m^2 - \psi_0^2\} + 
o\biggl({1\over N}\biggr)
\autoeq$$
with
$$
\psi_0={1\over 2} \biggl\{ {{J^2}\over {m^4}} + \dot G(m^2)\biggr\} 
\sqrt{\lambda},
\autoeq$$
from (\the\extrem),
and $m^2=r+\sqrt \lambda \psi_0$.  If the triple limit 
$N\rightarrow\infty, j 
\rightarrow \infty, u \rightarrow \infty$, such that $\lambda$ and $J$ 
are finite, is taken one finds
  
$$
\tilde w = {1 \over 2} \{G(m^2)-J^2/m^2 - \psi_0^2\} + o\biggl({1\over N}
\biggr)
$$
as the reduced free energy density for the model.
\bigskip
The effective potential, for constant $J$, is obtained from the 
Legendre transform with $\phi ={{\partial \tilde w}\over {\partial J}} = 
-{{J}\over{m^2}}$ so that 
$\phi =  \langle \varphi \rangle /\sqrt N$,
$$
\tilde\Gamma(\phi,r,\lambda) = {1 \over N}\Gamma(\phi,r,\lambda)=\tilde w 
- \phi J = {1\over 2}\{G(m^2)+m^2\phi^2-\psi_0^2\}+o \biggl({{1}\over{N}}
\biggr)
\autoeq$$
with $$\psi_0={1\over 2}\{\phi^2+\dot G(m^2)\}\sqrt\lambda. \autoeq$$
\bigskip
Eliminating $\psi_0$ using $\psi_0=(m^2-r)/\sqrt\lambda$ and re-arranging 
gives
$$\tilde\Gamma(\phi,r,\lambda) = {1\over 2}\biggl\{G(m^2) - m^2\dot G(m^2) 
+ {{m^4}\over{\lambda}} - {{r^2}\over{\lambda}}\biggr\} + o\biggl({1\over N}
\biggr)
\autoeq$$
$$
m^2 = r + {{\lambda}\over{2}} \dot G(m^2) + {{1}\over{2}}\lambda\phi^2
\autoeq$$
\newcount\effmass\effmass=\equno
which is the form of the effective action used in [\the\DenjoeChrisb].
\bigskip
The metric will be difficult to calculate using the co-ordinates 
$(\phi,r,\lambda)$ because $m$ is defined only implicitly through 
equation (\the\effmass).  It is easier to perform a second Legendre transform 
on the variable $r$ to a new variable,
$$
X := {1\over 2} \int d^Dx\langle\varphi^2\rangle = {{\partial\tilde\Gamma}
\over{\partial r}}\biggr\vert_{\phi,\lambda}, 
$$
and define 
$$\tilde\Xi(\phi,X,\lambda) = \tilde\Gamma-r{{\partial \tilde\Gamma}
\over{\partial r}}.
\autoeq$$
\bigskip
\noindent First note that (\the\effmass) gives
$$
{{\partial m^2}\over{\partial r}}\Vline\raise -12pt\hbox{$\phi,\lambda$} = 
{{1}\over{1-{\lambda\over 2}}\ddot G}
\autoeq$$
thus $X = (m^2-r)/\lambda$ or
$$
X = {{1}\over{2}}\{\dot G(m^2)+\phi^2\}
\autoeq$$
\newcount\Xeffmass\Xeffmass=\equno
using (\the\effmass) .  This gives
$$
\tilde\Xi(\phi,X,\lambda) = {1\over 2} \{G(m^2)-m^2 \dot G(m^2)+\lambda X^2\}
\autoeq$$
\newcount\Chit\Chit=\equno
with $m^2(\phi,X)$ given implicitly by (\the\Xeffmass).  Note that $m^2$ is 
independent of $\lambda$ when expressed as a function of $\phi$ and $X$
and it is this observation that simplifies the calculation of the
metric and and curvature when the $(\phi, X,\lambda)$ co-ordinate
system is used.
\bigskip
Using (\the\Xeffmass) we find 
$$
{{\partial m^2}\over{\partial X}}\biggl\vert_{\phi,\lambda} = {{2}\over 
{\ddot G}} \qquad\qquad
{{\partial m^2}\over{\partial \phi}}\Vline\raise -12pt\hbox{$X,\lambda$} 
= - {{2\phi}\over 
{\ddot G}}\qquad\qquad{{\partial m^2}\over{\partial \lambda}}
\Vline\raise -12pt\hbox{$\phi,\lambda$} 
=0 .
\autoeq$$
\newcount\dml\dml=\equno
The metric can now be determined using the results of appendix 1 for the 
Hessian of a Legendre transform,
$$
G_{ab} = N
\pmatrix 
{
{m^2-{{2\phi^2}\over{\ddot G}}}&{{2\phi}\over{\ddot G}}&0\cr
  {{2\phi}\over{\ddot G}} & {\lambda- {{2}\over{\ddot G}}} & 0\cr
 0&0&0\cr 
}+ o(1)
\autoeq$$
\newcount\degenmetric\degenmetric=\equno
i.e. the metric is degenerate at this order, since the Hessian of 
$\tilde\Xi$ has a zero mode in the $\lambda$-direction.  This 
degeneracy is lifted by including the $o(1/N)$ corrections to $\tilde\Xi$.
\bigskip
In order to determine the order one corrections to (\the\degenmetric)
one must calculate
the order one corrections to the partition function.
Consider therefore equations (\the\Zpsi) and (\the\Seff).
Including the order $\varepsilon^2$ 
terms from (\the\psioh) gives
$$\eqalign{
S_{eff} = {N \over 2} V& [G(m^2) - \psi_0^2 - J^2/m^2]\quad+\cr
{1\over 4}\int d^D&x\int d^Dy \;\varepsilon(x)\biggl[-\lambda{\delta(x-y)
\over (-\bigtriangledown_x^2+m^2)
(-\bigtriangledown_y^2+m^2)}
-2\delta(x-y)\biggl(1+{{\lambda J^2}\over{(m^2)^3}}\biggr)\biggr] 
\varepsilon(y)\cr
&\qquad\qquad\qquad\qquad\qquad\qquad+ o(\varepsilon/\sqrt N)^3.\cr}
\autoeq$$
After a contour
rotation, [\the\ZinnJustin], a Gaussian integral over $\varepsilon$ gives, up 
to an irrelevant constant independent of $J,r$ and $\lambda$,
$$
w = {N \over 2} \left\{G(m^2) - J^2/m^2 - \psi_0^2\right\} + 
{1 \over 2} \ln \det F + o\biggl({1\over N}\biggr)
\autoeq$$
\newcount\wone\wone=\equno
where $F$ is diagonal in momentum space,
$$
F(p) = 1+ {{\lambda J^2}\over{(m^2)^3}}  + {\lambda \over 2}
 \int {{d^Dq}\over{(2\pi)^D}}  {{1}\over{(p^2+m^2)}}  {{1}
\over{\left\{(p-q)^2+m^2\right\}}},
\autoeq$$
and
$$\ln \det F(p) = \int {{d^Dp}\over{(2\pi)^D}} \ln F(p) 
\equiv L(J,r,\lambda).\autoeq
$$
\newcount\Ldef\Ldef=\equno
The comments made after equations 
(\the\Gdef) - (\the\Gdddef) also apply here, the 
function $L$ 
depends on the geometry of D-dimensional space.  It is shown in appendix 1,
that the Legendre transform of any differentiable function of the form

$$
w(g) = w_0(g)+ {1\over 2N}L(g) + o\biggl( {1\over N^2} \biggr)
\autoeq$$
is

$$
\tilde\Gamma(\phi) =\tilde\Gamma_0(\phi)+{1\over 2N}L(g(\phi))
+ o\biggl( {1\over N^2} \biggr)
\autoeq$$\newcount\LT\LT=\equno
where $\phi = {{dw}\over{dg}}$ can be inverted to give 
$g(\phi)=g_0(\phi)+{1\over N}g_1(\phi)+o({1\over N^2})$  and \hfill\break
\hbox{$\tilde\Gamma_0(\phi) = w_0(g_0(\phi)) -g_0(\phi)\phi$}.
Thus the double Legendre transform of (\the\wone) is, 
using (\the\Xeffmass) and (\the\Chit)

$$\eqalign{
\tilde\Xi(\phi,X,\lambda) = &{1\over N}\Xi(\phi,X,\lambda)\cr
= &{1\over 2}\left\{ G(m^2) - m^2 \dot G(m^2) + \lambda X^2 \right\}
+ {{1}\over{2N}} \int {{d^Dp}\over{(2\pi)^D}} \ln F(p)
+ o\biggl({{1}\over{N^2}}\biggr)\cr}
\autoeq$$

\noindent with $m^2(\phi,X)$ determined implicitly by

$$
X = {1 \over 2} \left\{\dot G(m^2) + \phi^2\right\}
\autoeq
$$
and
$$
F(p) = 1 + {{\lambda \phi^2}\over{m^2}} + {\lambda \over 2}
 \int {{d^Dq}\over{(2\pi)^D}} {{1}\over{(q^2+m^2)}} {{1} \over 
{ \left\{ (p-q)^2 +(m^2) \right\} }}.
\autoeq$$
The function $F(p)$ encodes the order one corrections to the metric.
Expressing $L(p)$ in equation (\the\Ldef) 
as a function of $(\phi, X, \lambda)$ gives,
$$
G_{ab} = N
\pmatrix 
{
{m^2-{{2\phi^2}\over{\ddot G}}}&{{2\phi}\over{\ddot G}}&0\cr
  {{2\phi}\over{\ddot G}} & {\lambda -{{2}\over{\ddot G}}} & 0\cr
 0&0&0\cr 
}
+ {1 \over 2}
\pmatrix 
{
{{\partial^2L}\over{\partial\phi^2}}&{{\partial^2L}\over{\partial\phi\partial
 X}}&0\cr
  {{\partial^2L}\over{\partial\phi\partial X}} &  {{\partial^2L}\over
{\partial X^2}} & 0\cr
 0&0&-{{\partial^2L}\over{\partial\lambda^2}}\cr 
}
+ o\biggl( {{1}\over{N}}\biggr).
\autoeq$$
\newcount\metrix\metrix=\equno

\noindent Note that $G_{ab}$ is positive definite since $\ddot G<0$ 
and $L^{\prime\prime} 
= {{\partial^2L}\over{\partial\lambda^2}} < 0$.
\bigskip
The connection co-efficients can now be evaluated in a straightforward but 
tedious manner (remembering that the matrix is curl free in the chosen 
co-ordinate system \hbox{$G_{ab,c}=G_{ac,b}$} etc.) and they are enumerated in 
appendix 2.
\bigskip
The components of the Ricci tensor are
\bigskip
$$
{R^X}_X = {R^\lambda}_\lambda = - {{1}\over{2L^{\prime\prime}}} 
{(m^2{\ddot G}-2\phi^2)
\over\left\{\lambda(m^2{\ddot G}-2\phi^2)-
2m^2\right\}}
\left[ {L^{\prime\prime\prime}\over{L^{\prime\prime}} } 
+ {(m^2{\ddot G}-{{2\phi^2}})\over\left\{\lambda(m^2{\ddot G}-{{2\phi^2}})
-{{2m^2}}\right\}} \right] 
 + o\biggl( {{1}\over{N}} \biggr)
\autoeq$$\newcount\Ricciten\Ricciten=\equno
and ${R^a}_b = o (1/N)$ otherwise. As a reminder, a dot denotes ${{\partial}\over
{\partial m^2}}$ while a prime denotes ${{\partial}\over{\partial\lambda}}$.
\bigskip
The geometry is essentially such that all of the sectional curvature is in 
the $X-\lambda$ planes for constant $\phi$.  The sectional curvature in 
the $X-\phi$ and $\lambda-\phi$ planes is of order $1/N$, (see e.g. 
\autoref\newcount\Pauli\Pauli=\refno page 46).  
Thus, to this order, all of the geometry is encapsulated 
in the Ricci scalar, which is twice the Gaussian curvature of the surfaces 
of constant $\phi$,
$${\cal R}= - {{1}\over{L^{\prime\prime}}} 
{(m^2{\ddot G}-{2\phi^2})
\over\left\{\lambda(m^2{\ddot G}-{{2\phi^2}})-
{{2m^2}}\right\}}
\left[ {{L^{\prime\prime\prime}}\over{L^{\prime\prime}}} 
+ {(m^2{\ddot G}-{{2\phi^2}})\over\left\{\lambda(m^2{\ddot G}-{{2\phi^2}})
-{{2m^2}}\right\}} \right] 
 + o\biggl( {{1}\over{N}} \biggr)
\autoeq$$
\newcount\Ricci\Ricci=\equno
\medskip
\noindent with $\ddot G$ and $L$ given in equations (\the\Gdddef) and 
(\the\Ldef) and 
$m^2$ $(\phi, X)$ 
(the effective mass) defined implicitly in equation (\the\Xeffmass).  
\bigskip
Having obtained  ${\cal R}$
we can of course use whatever co-ordinate system we 
wish, and it is convenient now to change from $(\phi, X, \lambda)$ 
to $(\phi, m^2, \lambda)$. 
It would have been very tedious to have used $(\phi, m^2, \lambda)$
from the start because they are related to $(\phi, X, \lambda)$ non-linearly
and so neither equation (\the\Metric), nor its analogous form for
Legendre transformed variables, is valid in the $(\phi, m^2, \lambda)$
system.
The actual form 
of ${\cal R}(\phi,m^2,\lambda)$ now depends on
$$
\ddot G(m^2) = -\int {{d^Dp}\over{(2\pi)^D}}  {{1}\over{(p^2+m^2)^2}}
\autoeq$$\Gdddef=\equno
and
$$L = \int {{d^Dp}\over{(2\pi)^D}} \ln \biggl[ 1 + 
{{\lambda\phi^2}\over {m^2}} + {{\lambda}\over{2}} \int {{d^Dq}\over
{(2\pi)^D}} {{1}\over{(q^2+m^2)}} {{1}\over( {(p-q)^2+m^2}) } 
\biggl].
\autoeq$$
\newcount\Lint\Lint=\equno
\bigskip
The geometry of D-dimensional space has not yet been specified ---
it could be infinite Euclidean space or a D-dimensional torus are 
even a lattice with a finite set of points, in which case 
$\bigtriangledown^2$ is a matrix and $\int {{d^Dp}\over{(2\pi)^D}}
\rightarrow Trace$.
\bigskip
In the next section we shall examine the geometry for infinite 
3-dimensional Euclidean space, using a cut-off to regularise the 
integrals --- this case is of special interest, because the model is 
known to exhibit two fixed points --- one is the Gaussian fixed point leading 
to free field theory in the $UV$ direction and the other is the Wilson-Fisher
fixed point in the $IR$ direction, 
which is equivalent to the spherical model in the $N\rightarrow \infty$ 
limit [\the\Stanley].
\bigskip
\vfill\eject
\leftline{\bf \S 4 The O(N) Model in 3-Dimensions} 
\bigskip
The geometry on the space of couplings described by equation (\the\Ricci)
 will now be examined for the 
case of D=3, flat, Euclidean space.  The integral in (\the\Gdddef) is finite for 
D=3 and gives
$$
\ddot G (m^2) = -{1\over 2\pi^2}\int_0^\Lambda{p^2dp\over{(p^2+m^2)^2}} 
= -{1\over 4\pi^2}\Bigl\{{1\over m}\tan^{-1}\Bigl({\Lambda\over m}\Bigr)-
{\Lambda\over{\Lambda^2 + m^2}}\Bigr\}.
\autoeq$$
For simplicity the asymptotic form
$$\lim_{\Lambda\rightarrow\infty}\ddot G (m^2) =- {{1}\over{8\pi m}}\autoeq$$
\newcount\threedgdd\threedgdd=\equno
will be used below, but it should be borne in mind that the final
expression is only valid for ${m\over\Lambda}<<1$.
The $q$-integral in equation (\the\Lint) is also finite and gives
$$
L = {{1}\over {2\pi^2}}\int_0^\Lambda p^2dp\ln \biggl\{1 + {{\lambda\phi^2}
\over{m^2}} + {{\lambda}\over{8\pi p}} \tan^{-1} \biggl( {{p}\over{2m}}\biggr)
 \biggr\}
\autoeq
$$
where a cut-off, $\Lambda$, has been introduced because the $p$-integral 
diverges.  It is particularly easy to take derivatives of $L$ with respect 
to $\lambda$, the nth derivative being
$$
L^{(n)} =  {{(-1)^{n-1} (n-1)!}\over{2\pi ^2}} \int_0^\Lambda p^2dp\left[ 
{{{\phi^2p\over m^2} + {{1}\over{8\pi}} \tan ^{-1} \left({p\over 2m}\right)}
\over{p + {{\lambda\phi^2p}
\over{m^2}} + {{\lambda}\over{8\pi}} \tan^{-1} \left( {{p}\over{2m}} \right)
 }} \right]^n,
\autoeq
$$
\newcount\Ldiffn\Ldiffn=\equno
(note that ${\partial\over\partial\lambda}\bigl\vert_{\phi,X}=
{\partial\over\partial\lambda}\bigl\vert_{\phi,m^2}$ since 
${\partial m^2\over\partial\lambda}\bigl\vert_{\phi,X}=0$ from equation
(\the\dml)).

Using (\the\threedgdd) in (\the\Ricci) gives the Ricci scalar as
$$
{\cal R} = - {{ ({m\over 16\pi}  +  \phi^2)}\over{L^{(2)}} 
\left\{ {{\lambda}}
 ({m\over 16\pi} +\phi^2) + m^2 \right\} }\left[ {{L^{(3)}}
\over{L^{(2)}}} + {{({m\over 16\pi} +\phi^2)}\over \left\{ {{\lambda}}
 ({m\over 16\pi} +\phi^2) + m^2 \right\} }  \right]  + o(1/N)
\autoeq
$$
where $L^{(2)}$ and $L^{(3)}$ are defined in 
(\the\Ldiffn).  If the limit $\Lambda 
\rightarrow \infty$ is taken, $L^{(n)}\sim \Lambda^3 \Rightarrow 
{\cal{R}} \sim 1/\Lambda^3 \rightarrow 0$, but this is really throwing 
away important geometric information.  Better is to observe that 
${\cal {R}}$ has dimensions of mass$^{-3}$, $\phi^2$ and $\lambda$ 
both have dimensions of mass in 3-D so, following Zinn-Justin [\the\ZinnJustin], 
define dimensionless parameters 
$$
\bar\lambda = 
{{\lambda}\over 16\pi\Lambda} \quad , \quad \bar\phi^2 = 
{{16\pi\phi^2}\over{\Lambda}} \quad , \quad \bar m = {m\over\Lambda}
\qquad\hbox{and} \qquad \bar{\cal R} = \Lambda^3 {\cal {R}}.
\autoeq
$$
The last equation here is equivalent to a conformal rescaling 
by $\Lambda^{-3}$ which renders the metric dimensionless.
Now define 
$$
\bar L^{(n)} :=(16\pi)^n {L^{(n)}\over \Lambda^{3-n} } 
= {(-1)^{n-1}(n-1)!(16\pi)^n\over
{2\pi^2}} \int_0^1 z^2dz  
\Biggl[ {\bar\phi z + 2\bar m^2\tan^{-1}
(z/2\bar m)\over \bar m^2z+\bar\lambda\bar\phi z+2\bar\lambda\bar m^2
\tan^{-1}({z/2\bar m}) } \Biggr]^n,\autoeq
$$
\newcount\Lbarn\Lbarn=\equno
in terms of which the rescaled Ricci scalar is 
$$
{\bar{\cal {R}} } = -{1\over\bar L^{(2)}}
{ (\bar m + \bar\phi^2)\over 
\{\bar m^2 + \bar\lambda(\bar m + \bar\phi^2)\}} 
\Biggl[ {\bar L^{(3)}\over\bar L^{(2)}} + 
{(\bar m + 
\bar\phi^2)\over \{\bar m ^2 + \bar\lambda  (\bar m +\bar\phi^2)\} }
  \Biggr]  + o(1/N)
\autoeq
$$
\newcount\RBar\RBar=\count90
which is finite even when $\Lambda\rightarrow\infty$, provided
$\bar\phi$, $\bar m$ and $\bar\lambda$ are kept finite and
are not all zero.   
The Ricci scalar is shown in figures 1---4, where 
it is graphed as a function of $\bar m$ and $\bar\phi^2$ for 
four values of $\bar\lambda$, $
\bar\lambda=0.1$, $\bar\lambda=0.2$, $\bar\lambda=0.3$ and 
$\bar \lambda=5.0$.  
In order to produce these graphs, the integrals in (\the\Lbarn) were performed 
numerically.
\bigskip
The Ricci scalar is infinite at $\bar\lambda=\bar m=\bar\phi^2=0$, 
corresponding to the Gaussian fixed point, but is finite  elsewhere.  
If either of the two 
variables, $\bar\phi, \,\, \hbox{or}\,\, \bar\lambda$ becomes 
large, then $\bar{\cal R}$ tends to a negative constant.
$$
\bar{\cal R}\vert_{\bar\lambda\rightarrow\infty}=
\bar{\cal R}\vert_{\bar\phi\rightarrow\infty}=-6\pi^2
+o\biggl( {{1}\over{N}}\biggr).
\autoeq
$$
The limit for large $\bar m$ can be obtained from (\the\Ricci), (\the\Gdddef)
and (\the\Lint)  directly, by taking this limit before performing the 
integrals, avoiding
the constraint $\bar m<<1$. One finds again
$$\bar{\cal R}\vert_{\bar m\rightarrow\infty}= -6\pi^2
+o\biggl( {{1}\over{N}}\biggr),\autoeq$$
which is the same value as one gets by n\"aively putting 
$\bar m\rightarrow\infty$ in equation (\the\RBar).

Of particular interest is the curvature along the critical 
line. Setting $\bar\phi=0$ first and then letting  $\bar m\rightarrow 0$
gives
$$\eqalign{
\bar{\cal R}\vert_{\bar\phi=0,\bar m\rightarrow \;0} &=
\left({2\over\bar\lambda^2 }\right)
{\left\{1+6\pi\bar\lambda +4(\pi\bar\lambda)^2-4\pi\bar\lambda 
(1+\pi\bar\lambda )^2
\ln\left(1+{1\over\pi\bar\lambda }\right)\right\}
\over 
\left\{1+2\pi\bar\lambda -2\pi\bar\lambda(1+\pi\bar\lambda )
\ln\left(1+{1\over\pi\bar\lambda }\right)\right\}^2}\cr
&\longrightarrow\cases{ 
-6\pi^2&\hbox{$\bar\lambda\rightarrow\infty$}  \cr
{2/\bar\lambda^2}&\hbox{$\bar\lambda\rightarrow 0$} \cr}\cr}
\autoeq
$$\newcount\Rcritline\Rcritline=\equno
which is shown in figure 5.  There is actually a discontinuity in 
$\bar{\cal R}$ 
across this line, if we keep $\bar m = 0$, which is the critical line in 
the $\bar \phi - \bar \lambda$ plane.  For $\bar \phi \neq 0$ and  
$\bar m = 0$, $\bar{\cal R}= 0$ $\forall \;\bar\lambda$, but for 
$\bar\phi = 0$ and $ \bar m \rightarrow 0$ , $\bar{\cal R} \neq 0$, 
(except at one value of 
$\bar \lambda\approx 0.2$). This behaviour is shown in figure 6,
where $\bar{\cal R}$ is graphed as a function of $\bar\phi$ for
$\bar m=0$ and a generic value of $\bar\lambda$.
The discontinuity in $\bar{\cal R}$ is due to the non-analyticity 
of the specific heat at the critical point. As explained in the 
introduction, the metric is defined in terms of second derivatives 
of the free energy and the Riemann tensor involves third derivatives 
of the reduced free energy with respect to any one parameter 
(e.g. temperature), hence one 
naively expects the Ricci scalar to diverge at a critical point.  
This does not happen here, except at the Gaussian fixed point 
$\bar\lambda = 0$, because the specific heat exponent $\alpha = -1
+o(1/N)$ is negative for the O(N) model in 3-dimensions --- the Ricci
 scalar is finite, but discontinuous, i.e. its derivative diverges at 
the critical line.
\bigskip
Let us examine the critical line, $\bar m = 0$, more closely for a 
fixed value of $\bar \lambda$.  In 3-dimensions $\dot G(m^2)$ diverges, 
so introducing a cut-off, equation (\the\Gddef) yields
$$
\dot G(m^2) = - {{m}\over{4\pi}} - {{\Lambda}\over{2\pi^2}}
\autoeq
$$
so $m = - 4\pi\{\dot G(m^2) - \dot G(0)\}$.  Equation 
(\the\effmass) can now be 
solved to give $m(r, \phi, \lambda)$, or in terms of dimensionless 
quantities
$$
\bar m = \sqrt{{{\bar\lambda^2+{\bar\lambda\bar\phi^2\over 2}+t}}} - 
\bar\lambda
\autoeq
$$ 
where $t = \biggl(r+{{\lambda\dot G(0)}\over{2}}\biggr)/\Lambda^2$ 
is the reduced temperature.  The addition of ${{\lambda\dot G(0)}\over{2}}$ 
to the bare parameter, $r$, is the usual mass shift.  The critical 
temperature $t=0$, gives $\bar m=0$ for $\bar\phi=0$ 
(vanishing external field) 
but for $t<0$, $\bar m=0$ for $\bar\phi=4|t|/ \bar\lambda$, which is the 
critical line.  Along a line specified by 
$\bar\phi\neq 0$, $\bar m=0$ and a fixed value of $\bar\lambda$, 
the specific heat 
is finite [\the\Wallace] --- the 
line along which it diverges lies in the unstable
region and is known as the pseudo-spinodal 
line, this only coincides with the critical line, $\bar m=0$, 
for $\bar \phi=0$.  
\bigskip
Finally, let us consider renormalisation group flow.  Following 
Zinn-Justin [\the\ZinnJustin], 
define $\beta$-functions for the three parameters $\bar\phi, t, \bar\lambda$ by
$$
\beta^{\bar\phi} = \Lambda {{d\bar\phi}\over{d\bar\Lambda}} 
= - {1\over 2}\bar\phi
$$
$$
\beta^t = \Lambda {{dt}\over{d\bar\Lambda}} = - 2t
$$
$$
\beta^{\bar\lambda} = \Lambda {{d\bar\lambda}\over{d\lambda}} = - \bar\lambda
\autoeq
$$
These are simply the canonical dimensions since $(\bar\phi,t,\bar\lambda)$ 
are bare parameters, which are finite for finite cut-off $\Lambda$.  In 
terms of the variables $\phi$, $X$ and $\lambda$, let $\bar X = X/ \Lambda$ 
be dimensionless and then
$$
\beta^{\bar\phi} = -{1\over 2}\bar\phi \qquad \beta^{\bar X} = -\bar X \qquad 
\beta^{\bar \lambda} = -\bar\lambda
\autoeq
$$
These represent a vector flow on the space of parameters and we shall now 
investigate the dynamics of this vector flow, in particular we can ask: 
how is this flow related to geodesics of the metric (\the\metrix)? 
\bigskip
For any curve $\bar\phi(\Lambda),\bar X(\Lambda),\bar
\lambda(\Lambda)$ parameterised by $\Lambda$, the geodesic equation is
$$
{{d^2x^\mu}\over{d\Lambda^2}} + \Gamma_{\rho\sigma}^\mu  {{dx^\rho}\over{d
\Lambda}}  {{dx^\sigma}\over{d\Lambda}} = - cx^\mu
\autoeq
$$\newcount\geodeq\geodeq=\equno
where $c(x)$ is a function which allows for the possibility that $\Lambda$ 
might not be an affine parameter, 
\autoref\newcount\EllisHawking\EllisHawking=\refno. Using the connection 
co-efficients in appendix 3, one finds that the condition that both the 
order N and the order one contributions satisfy equation (\the\geodeq), with 
$\bar\phi,\bar X$ and $\bar\lambda$ all order one, is very restrictive 
and the only solution is $\bar X=\bar\phi = 0$, though $\bar\lambda$ 
can be non-zero, provided $c$ is a function of $\bar\lambda$ alone, given by
$$
c(\bar\lambda)= 1 + {{\bar\lambda\bar L^{(3)}}\over{2\bar L^{(2)}}}.
\autoeq
$$
This leads to the somewhat surprising result that the line of crossover 
from the Wilson-Fisher fixed point to the Gaussian fixed point is a 
geodesic, but none of the other RG trajectories is.  The physical 
significance of this result will be examined in the next section.

\vfill\eject

{\leftline {{\bf \S 5 Relative Entropy}}}
\bigskip
In this section a physical interpretation of the geodesic flow, unveiled 
in the previous section, is given.  The metric used in the previous analysis 
is related to the concept of relative entropy in statistical mechanics 
(for a consideration of relative entropy in field theory, see 
\autoref\newcount\DenjoeHose\DenjoeHose=\refno).
For a 
discrete probability distribution $p_i(g)$, $i=1,\ldots,r$, 
depending on some set of 
parameters $g^a$, $a=1,\dots,n$, the relative entropy of $g^a$ relative to 
$g^{a^\prime}$ is defined to be ,
\autoref\newcount\Kullback\Kullback=\refno,
$$
{\cal S}_R (g, g^\prime) = - \sum_{i=1}^{r} p_i(g) \ln 
\left\{p_i(g)/p_i(g^\prime)\right\}.
\autoeq$$
For a continuous probability distribution, the discrete sum becomes 
a functional integral with
$$
p_i(g) \rightarrow {e^{-S[\varphi, g]}\over{Z(g)}}
\autoeq$$\newcount\probmeasure\probmeasure=\equno
so
$$
{\cal S}_R(g,g^\prime) = <S(g)>_g - <S(g^\prime)>_g + W(g^\prime) - W(g)
\autoeq$$
where $W(g) = - \ln Z(g)$, and all expectation values use the measure 
appropriate to $g$ as in equation (\the\probmeasure), not $g^\prime$.
\bigskip
Dividing by the volume of D-dimensional space, so as to work with specific 
quantities, one defines the relative entropy per unit volume to be
$$
s_R(g,g^\prime) = - {{1}\over{V}} \bigl\{<S(g^\prime)>_g - <S(g)>_g\bigr\} 
+ w(g^\prime)-w(g).
\autoeq$$
\newcount\specent\specent=\equno
If $g^{a^\prime} = g^a+\delta g^a$, with
$\delta g^a$ small, we have 
$$
\eqalign{
S(g^\prime) &= S(g) + \delta 
g^a \partial_aS(g) + {1\over 2}\delta g^a\delta g^b \partial_a\partial_b S(g) + 
\dots \cr w(g^\prime) &= w(g) + \delta g^a \partial_a w(g) + {1\over 2} \
\delta g^a \delta g^b \partial_a\partial_b w(g) +\dots }\autoeq$$
and the terms linear in $\delta g$ cancel in (\the\specent), 
since $\partial_aw={{1}\over {V}}
<\partial_a S>$, giving 
$$
-s_R(g,g+\delta g) = {1\over 2} \biggl\{ {1\over V} <\partial_a\partial_b S>
 - \partial_a\partial_bw\biggr\} \delta g^a\delta g^b + o(\delta g)^3.
\autoeq$$
Thus the metric defined in 
(\the\metricb) is completely equivalent to the infinitesimal
relative entropy, and the distance between two points $g_A$ and $g_B$ along 
a curve $g(\tau)$, parameterised by $\tau$, is given by
$$
d = \int_{\tau_A}^{\tau_B} \sqrt{-2s_R} \;dt = 
\int_{\tau_A}^{\tau_B} \sqrt{G_{ab}\dot g^a \dot g^b} d\tau,
\autoeq$$
where $\dot g^a={dg^a\over d\tau}$.
Note that $S_R(g, g^\prime)\neq S_R(g^\prime,g)$ for finite $g^\prime - g$ so
$S_R$ itself cannot be interpreted as a distance function. 

The conclusions of 
the previous section can now be rephrased by saying 
that, in the large $N$ limit 
of the $O(N)$ model in 3-dimensions, the line of crossover between the 
Wilson-Fisher fixed point and the Gaussian fixed point is a line of 
extremal relative entropy.  At least for the segment of the line along 
which $\bar{\cal R}<0$, i.e.
$$
\bar\lambda_0\approx 0.2 <\bar\lambda< \infty\autoeq
$$
where $\bar\lambda_0$ is the value of $\bar\lambda$ at which 
$\bar{\cal R}=0$ (figure 5), 
one can be 
confident that the relative entropy is maximised, since there can be no 
conjugate points for $\bar{\cal R}<0$, [24].
\bigskip 
{\leftline{{\bf \S 6 Conclusions}}}
\bigskip
The concept of a geometry on the space of couplings, and its relation to 
the vector flow of the renormalisation group equation, has been 
investigated in the particular case of the O(N) model in the limit of 
$N \rightarrow \infty$.  The space of couplings in this case is three 
dimensional and can be parameterised by the vacuum expectation of the 
field, a mass and the $\varphi^4$ coupling.  The metric adopted, 
$$
G_{ab} = \int d^Dx \langle\tilde\Phi_a(x)\tilde\Phi_b(0)\rangle,
\autoeq$$
is the matrix given by taking the zero momentum 
limit of two point correlators of the composite operators associated 
with the couplings, which ought to capture the infra-red behaviour of 
the theory, but would not be expected to give useful information in the 
ultra-violet. This is borne out when the curvature is calculated and in 
D=3, in the infinite volume limit,
the Ricci scalar is found to diverge at the Gaussian (ultra-violet) 
fixed point, but is finite (and negative) at the Wilson-Fisher (infra-red) 
fixed point.  This statement is independent of the co-ordinate system 
used --- it does not matter whether one uses bare or renormalised couplings.  
This is not true of the components of the metric ---  divergences in the 
metric could be due to either genuine singularities of the geometry or 
could be merely co-ordinate artifacts due, for example, to a
parameterisation using bare couplings.
\bigskip
In particular, the Ricci scalar is a smooth 
monotonically increasing function 
along the RG trajectory between the Wilson-Fisher and the Gaussian fixed 
points, although it is not differentiable in one of the directions 
transverse to the line of cross-over --- 
reflecting the fundamental non-analyticity of the free energy at
the critical line.
\bigskip
It was also noted in section 4 that this cross-over line is a geodesic in 
the geometry described here but none of the other RG trajectories, which 
miss the Gaussian fixed point,
is a geodesic. This property is equivalent 
to the statement that the relative entropy is maximised along this curve.
The geodesic nature of some renormalisation group trajectories in simpler 
models was noted in [\the\geodrg], and would seem to hint at a possible 
variational formalism for the RG, but this requires further study.
\bigskip

The author wishes to thank Denjoe O'Connor for many  discussions
about the renormalisation group and relative entropy. 
This work received partial financial
support from an exchange fellowship between the Royal Irish Academy and
the Royal Society and it is a pleasure to thank the Department of Physics
and Astronomy at the University of Edinburgh, Scotland, 
whose hospitality allowed the completion of the manuscript. The work
also received sponsorship from
Baker Consultants Ltd., Ireland, networking specialists
(http://www.baker.ie).
\vfill\eject
\leftline {\bf {Appendix 1 - Legendre Transforms}}
\bigskip

Consider a differentiable function $w(g)$ of $n$ variables  
$g^a, a=1,\dots,n$, and the corresponding Hessian
$$
G_{ab} = - {{\partial^2w}\over{\partial g^a\partial g^b}} \equiv -w_{ab}.
\autoeq$$
If one changes variables to $(g^{a^\prime})=(\phi_1,g^2,\dots,g^n)$, where
$$
\phi_1(g) = {{\partial w}\over{\partial g^1}}
\autoeq$$
is the Legendre transform variable 
of $g^1$, then the corresponding co-ordinate 
transformation matrix is
$$
\partial g^{a^\prime} / \partial g^b = 
\pmatrix{
{w_{11}}&{w_{12}}& \dots & {w_{1n}}\cr
0&1&\dots&0\cr
\vdots & \vdots&\vdots&\vdots\cr
0&0&\dots&1 \cr }
\autoeq$$
and the inverse matrix is
$$
\partial g^b / \partial g^{a^\prime} = 
\pmatrix{
{{1}\over {w_{11}}}&-{{w_{12}}\over {w_{11}}}&\dots &-{{w_{1n}}
\over{w_{11}}}\cr
0&1&\dots&0\cr
\vdots&\vdots&\vdots&\vdots\cr
0&0&\dots&1 \cr }.
\autoeq$$\newcount\partialg\partialg=\equno
\bigskip
Treating $G_{ab}$ as a tensor (which requires endowing the original 
co-ordinate system, $g^a$, with a very special status as explained in 
the introduction) one finds
$$
\eqalign{
G_{a^\prime b^\prime}&={\Biggl[ \biggl({{\partial g}\over{\partial 
g^\prime}}\biggr)^T\Biggr]_{a\prime}}^{b}\;   G_{ab}\,{\biggl( 
{{\partial g}\over{\partial g^\prime}} \biggr)^b}_{b^\prime}\cr
&=\pmatrix{
\tilde\Gamma_{11}&0&\dots&0\cr
0&-\tilde\Gamma_{22}&\dots&-\tilde\Gamma_{2n}\cr
\vdots&\vdots&\vdots&\vdots\cr
0&-\tilde\Gamma_{2n}&\dots&-\tilde\Gamma_{nn}\cr }} 
\autoeq$$
where $\tilde\Gamma(\phi_1, g^2, \dots , g^n) = \{w(g)- \phi_
1 g^1\} 
\vert_{\phi_1=\partial w/\partial g^1}$ is the Legendre transform of 
$w$,\hfill\break
\hbox{$\tilde\Gamma_{11} = {{\partial^2\tilde\Gamma}\over{\partial(\phi_1)^2}} 
= -{{1}\over{w_{11}}}$} and $\tilde\Gamma_{\bar a\bar b} = 
{{\partial^2\tilde\Gamma}\over{\partial g^{\bar a}\partial g^{\bar b}}}$
with $\bar a,\bar b=2,\dots n$.
\bigskip
In deriving this result it is important to remember that, for $\bar a = 2, 
\dots , n$
$$
{{\partial\tilde\Gamma}\over{\partial g ^{\bar a}}}\bigg\vert_{\phi_1} =
 {{\partial w}\over{\partial g ^{\bar a}}}\bigg\vert_{g^1} + {{\partial w}
\over{\partial g^1}} \cdot {{\partial g^1}\over{
\partial g ^{\bar a}}} 
\bigg\vert_{\phi_1} - \phi_1 {{\partial g^1}\over{\partial g^{\bar a}}}
\bigg\vert_{\phi_1}
= {{\partial w}\over {\partial g^{\bar a}}}\bigg\vert_{g^1}  \autoeq$$
where $g^1(\phi_1, g^2, \dots, g^n)$ is determined by inverting the 
function $\phi_1(g^1,\ldots,g^n)
= \partial w(g)/\partial g^1$. Similarly
$$
{{\partial^2\tilde\Gamma}\over{\partial g ^{\bar a}\partial g^{\bar b}}}
\bigg\vert_{\phi_1} = {{\partial^2w}\over{\partial g ^{\bar a}\partial 
g^{\bar b}}}\bigg\vert_{g^1} + {{\partial ^2w}\over{\partial g^{\bar a}
\partial g^1}} {{\partial g^1}\over{\partial g ^{\bar b}}} 
\bigg\vert_{\phi_1}  
= w_{\bar a \bar b} - {{w_{1 \bar a} w_{1\bar b}}\over{w_{11}}} 
\autoeq$$
since ${{\partial g^1}\over{\partial g^{\bar b}}} = - {{w_{1\bar b}}\over 
{w_{11}}}$ from (\the\partialg). 
\bigskip
Alternatively, since the complete Legendre transform, $\Psi(\phi)$, 
where $\phi_1={{\partial w}\over{\partial g^1}}$,\break 
$\phi_2={{\partial 
\tilde\Gamma}\over{\partial g^2}}\vline_{\phi_1} = {{\partial w}
\over{\partial g^2}}\vline_{g^1}$ etc., has the property that the matrix
$$
{{\partial^2\Psi(\phi)}\over{\partial \phi_a\partial \phi_b}} \equiv 
\Psi^{ab}
\autoeq$$
is the inverse of ${{\partial^2w}\over{\partial g^a\partial g^b}}$, one has
$$
ds^2 = G_{ab}dg^adg^b = \Psi^{ab} d\phi_ad\phi_b
\autoeq$$
\bigskip
The  same argument can now be applied to the Legendre transform of 
$\Psi(\phi)$ in 
one variable, $\tilde\Xi(\phi_1, \dots , \phi_{n-1}, g^n) = \Psi(\phi) - 
\phi_ng^n$, to deduce that, in the co-ordinate system 
$\phi^{\hat a}=(\phi_1,\dots,\phi_{n-1}, g^n)$,
$$G_{\hat a\hat b}=
\left(
\matrix{
-\tilde\Xi_{11}&\dots& -\tilde\Xi_{1,{n-1}} & 0\cr
\vdots & \vdots&\vdots&\vdots\cr
-\tilde\Xi_{1,{n-1}}&\dots&-\tilde\Xi_{n-1, n-1}&0\cr
0&\dots&0&\tilde\Xi_{nn} \cr }
\right) \autoeq$$
and this is the form of the metric used in the text, with $n=3$ and 
$\phi^{\hat 1}=\phi$, $\phi^{\hat 2}=X$, $g^3=\phi^{\hat 3}=\lambda$.
\bigskip
Finally, a proof will be given of equation (\the\LT) in the text.  For 
simplicity we consider a function $w(g)$ of only one argument, but the 
results apply equally well to a function of more than one variable.  Let
$$
w(g) = w_0(g) + {{1}\over{2N}} L(g) + o\biggl( {{1}\over {N^2}} \biggr)
\autoeq$$
where $w_0$ and $L$ are independent functions of $g$, and $N$ is some large 
parameter.  The Legendre transform variable is
$$
\eqalign{
\phi (g) &= {\partial w_0\over\partial g} + {1\over 2N} 
{\partial L\over\partial g} +o\biggl( {1\over N^2}\biggr) \cr 
&= \phi_0(g) + {1\over N} \phi_1(g) + o\biggl( {1\over N^2}\biggr) \cr}
\autoeq$$\newcount\GT\GT=\equno
and
$$
\tilde\Gamma (\phi) = w_0(g) + {{1}\over{2N}} L(g) - \phi.g + 
o\biggl( {{1}\over{N^2}}\biggr)
\autoeq$$
Inverting equation (\the\GT), one obtains
$$
g(\phi) = g_0(\phi)+ {{1}\over {N}} g_1(\phi) + o\biggl( {{1}\over{N^2}}
\biggr)
\autoeq$$
where $g_0(\phi)$ and $g_1(\phi)$ are functions of $\phi$.
Therefore, Taylor expanding $w_0(g)$ and $L(g)$,
$$
\eqalign{
w_0(g) &= w_0(g_0) + {{1}\over{N}} g_1. {{\partial w_0}\over {\partial g}} 
\bigg\vert_{g_0} + o\biggl( {{1}\over{N^2}}\biggr) \cr
 L(g) &= L(g_0) + o\biggl( {{1}\over{N}}\biggr)}
\autoeq$$
leads to
$$
\eqalign{
\tilde\Gamma(\phi) & = w_0(g_0) + {{1}\over{N}} g_1 {{\partial w_0}
\over{\partial g}}\bigg\vert_{g_0} + {{1}\over{2N}} L (g_0) 
-\biggl( g_0(\phi) + {{1}\over{N}} g_1(\phi)\biggr)\phi
+ o\biggl({{1}\over{N^2}}\biggr) \cr 
&= w_0(g_0(\phi)) + {{1}\over{2N}} L(g_0(\phi)) - \phi . g_0(\phi) + 
o\biggl( {{1}\over{N^2}} \biggr) \cr }\autoeq$$
since
$$
\phi = {{\partial w_0}\over{\partial g}}\bigg\vert_{g_0} + o\biggl( {{1}
\over{N}} \biggr).
$$
Hence
$$\eqalign{
\tilde\Gamma(\phi) &= \tilde\Gamma_0(\phi) + {{1}\over{2N}} L(g_0(\phi)) + 
o\biggl( {{1}\over{N^2}} \biggr)\cr
&= \tilde\Gamma_0(\phi) + {{1}\over{2N}} L(g(\phi)) + 
o\biggl( {{1}\over{N^2}} \biggr)\cr}
\autoeq$$
where
$$
\tilde\Gamma_0(\phi) = w_0(g_0) - \phi . g_0
$$
which is equation (\the\LT).
\bigskip
{\leftline{{\bf Appendix 2 - Connection Co-efficients for O(N) model}}}
\bigskip
The metric is given in equation (\the\metrix) in the $(\phi, X, \lambda)$ 
co-ordinate system.
$$
G_{ab} =\pmatrix{
\gamma_{ij} & 0\cr 0 & 0 \cr} + {{1}\over{2}} \pmatrix{
L_{\phi\phi}& L_{\phi X} & 0 \cr L_{\phi X}& L_{XX} & 0 \cr 0 & 0 & -
L_{\lambda\lambda}\cr}  +o\biggl( {{1}\over{N}} \biggr)
$$
where $\gamma_{ij}$ is the 2 x 2 matrix
$$
\gamma_{ij} = {N} 
\pmatrix{ m^2-{{2\phi^2}\over{\ddot G}}&{{2\phi}
\over{\ddot G}}\cr
{{2\phi}\over{\ddot G}}& \lambda -{{2}\over{\ddot G}}\cr } $$
and $L_{\phi\phi} = {{\partial^2L}\over{\partial \phi^2}}$ etc.  
The functions $\ddot G$ and $L$ are given in equations (\the\Gdddef) 
and (\the\Lint) and 
the effective mass $m^2(\phi, X)$ is defined implicitly in equation 
(\the\Xeffmass).  
The inverse metric is
$$
G^{ab}= \pmatrix{ \gamma^{ij} & 0\cr 
0&-{{2}\over{L_{\lambda\lambda}}}   +o\biggl( {{1}\over{N}} \biggr) \cr} 
$$
with
$$
\gamma^
{ij} = {{1}\over{N\det\gamma}} \pmatrix{ \lambda -{{2}\over{\ddot G}}&
-{{2\phi}\over{\ddot G}}\cr
-{{2\phi}\over{\ddot G}}& m^2-{{2\phi^2}\over{\ddot G}}\cr }  +
o\biggl( {{1}\over{N^2}} \biggr). $$
\bigskip
The evaluation of the connection co-efficients is simplified by the 
observation that, in the co-ordinate system used here, the metric is 
curl free
$$
G_{ab,c} = G_{cb,a} = G_{ca,b} : = G_{abc},
$$
thus
$$
\Gamma_{bc}^{a} = {{1}\over{2}} G^{ad} G_{dbc}.
$$
Explicitly one finds, 
$$
\Gamma_{XX}^{X} = {{2m^2\dddot G}\over{\det \gamma(\ddot G)^3}} + 
o\biggl( {{1}\over{N}} \biggr)
$$

$$\Gamma_{X\phi}^{X} = -{{2\phi}\over{\det \gamma(\ddot G)^3}} 
(m^2\dddot G + \ddot G) 
+ o\biggl( {{1}\over{N}} \biggr)
$$

$$\Gamma_{\phi\phi}^{X} = -{{1}\over{\det \gamma(\ddot G)^3}} 
\biggl\{m^2(\ddot G)^2 + \phi^2 (2m^2\dddot G + 4 \ddot G) \biggr\}  + 
o\biggl( {{1}\over{N}} \biggr)
$$

$$
\Gamma_{XX}^{\phi} = - {{2\lambda\phi\dddot G}\over{\det \gamma
(\ddot G)^3}} 
+ o\biggl( {{1}\over{N}} \biggr)$$

$$\Gamma_{X\phi}^{\phi} = {{1}\over{\det \gamma(\ddot G)^3}} 
\biggl\{\lambda(\ddot G)^2-2\ddot G + 2\lambda\phi^2\dddot G\biggr\} 
+ o\biggl( {{1}\over{N}} \biggr)
$$

$$\Gamma_{\phi\phi}^{\phi} = {{\phi}\over{\det \gamma(\ddot G)^3}} 
\biggl\{4\ddot G - 3\lambda(\ddot G)^2 - 2\lambda\phi^2 \dddot G 
\biggr\} + o\biggl( {{1}\over{N}} \biggr)
$$

$$
\Gamma_{X\lambda}^{X} = {{1}\over{2\det \gamma}} \left(m^2-{2{\phi^2}\over
{\ddot G}}\right) + o \biggl( {{1}\over{N}} \biggr)
$$

$$
\Gamma_{X\lambda}^{\phi} =  - {{\phi}\over{\det \gamma \;\ddot G}} + 
o \biggl( {{1}\over{N}} \biggr)
$$

$$
\Gamma_{\phi\lambda}^{X} = o\biggl( {{1}\over{N}} \biggr) \quad \quad 
\Gamma_{\phi\lambda}^{\phi} = o \biggl( {{1}\over{N}} \biggr) 
$$

$$
\Gamma_{XX}^{\lambda} = {{N}\over{L_{\lambda\lambda}}} + o (1) \quad 
\quad \Gamma_{X\phi}^{\lambda} = {{L_{X\phi\lambda}}\over{2L_{\lambda
\lambda}}} + o\biggl( {{1}\over{N}} \biggr)$$

$$
\Gamma_{\phi\phi}^{\lambda} = {{L_{\phi\phi\lambda}}\over{2L_{\lambda
\lambda}}} + o\biggl( {{1}\over{N}} \biggr) \quad \quad  
\Gamma_{X\lambda}^{\lambda} = {{L_{X\lambda\lambda}}\over{2L_{\lambda
\lambda}}} + o\biggl( {{1}\over{N}} \biggr)$$
$$
\Gamma_{\phi\lambda}^{\lambda} = {{1}\over{2}} {{L_{\phi\lambda
\lambda}}\over{L_{\lambda\lambda}}} + o\biggl( {{1}\over{N}} \biggr) 
\quad \quad \Gamma_{\lambda\lambda}^{X} = o \biggl( {{1}\over{N}} \biggr)$$
$$
\Gamma_{\lambda\lambda}^{\phi} =  o\biggl( {{1}\over{N}} \biggr) 
\quad \quad \Gamma_{\lambda\lambda}^{\lambda} =  {{L_{\lambda\lambda\lambda}
\over{2L_{\lambda\lambda}}}} + o \biggl( {{1}\over{N}} \biggr),$$
where $L_{\lambda\lambda}={\partial^2 L\over \partial\lambda\partial\lambda}$
etc.
Using these expressions, the components of the Ricci tensor in equation 
(\the\Ricciten
) can be verified.
\vfill\eject
{\bf References}\hfill
\vskip .5cm
\bigskip
\item{[\the\Zam]} A.B Zamolodchikov, Pis'ma Zh. Eksp. Teor. Fiz. 
{\bf 43} (1986) 565
\smallskip
\item{[\the\DenjoeChris]} D. O'Connor and C.R. Stephens {\it Geometry, The Renormalisation Group
And Gravity}\break
in {\it Directions In General Relativity},\hfill\break
Ed. B.L. Hu, M.P. Ryan Jr., and C.V. Vishveshwava\hfill\break
Proceedings of the 1993 International Symposium,
Maryland, Vol 1, C.U.P. (1993)
\smallskip
\item{[\the\Fisher]} R.A. Fisher, Proc. Cam. Phil. Soc. {\bf 122} (1925) 700
\smallskip
\item{[\the\Rao]} C.R. Rao, Bull. Calcutta Math. Soc. {\bf 37} (1945) 81
\smallskip
\item{[\the\Amari]} S. Amari, {\it Differential Geometric Methods
in Statistics}, Lecture notes in Statistics {\bf 28} (1985) Springer
\smallskip
\item{[\the\Ruppeiner]} G. Ruppeiner, Rev. Mod. Phys. {\bf 67} (1995) 605
\smallskip
\item{[\the\Weinhold]} F. Weinhold, J. Chem. Phys. {\bf 63} (1975)
2479, {\bf 63} (1975) 2484, {\bf 63} (1975) 2488,\hfill\break
{\bf 63} (1975) 2496, {\bf 65} (1976) 559
\smallskip
\item{[\the\PV]} J.P. Provost and G. Vallee, Comm. Math. Phys. {\bf 76} (1980) 289
\smallskip
\item{[\the\ctheorem]} J.L. Cardy, Phys. Lett. {\bf 215B} 
(1988) 749\hfill\break
H. Osborn, Phys. Lett. {\bf 222B} (1989) 97\hfill\break
I. Jack and H. Osborn, Nuc. Phys. {\bf B343} (1990) 647\hfill\break
G.M. Shore, Phys. Lett. {\bf 253B} (1991) 380, {\bf 256B} (1991) 407
\hfill\break
A. Cappelli, J.I. Latorre and X. Vilasis-Cardona, 
Nuc. Phys. {\bf B376} (1992) 510\hfill\break
A.H. Castro-Neto and E. Fradkin, Nuc. Phys. {\bf B400} (1993) 525\hfill\break
F. Bastianelli, Phys. Lett. {\bf 369B} (1996) 249
\smallskip
\item{[\the\GeomRG]} B.P. Dolan, Int. J. Mod. Phys. {\bf A 9} (1994) 1261
\smallskip
\item{[\the\Sonoda]} H. Sonoda, {\sl Connections on Theory Space}
pre-print UCLA/93/TEP/21\hfill\break
K. Ranganathan, Nuc. Phys. {\bf B408} (1993) 180\hfill\break
K. Ranganathan, H. Sonoda and B. Zwiebach, Nuc. Phys. {\bf B414} (1994) 405
\hfill\break
B.P. Dolan, Int. J. of Mod. Phys. {\bf A10} (1994) 2439
\smallskip
\item{[\the\geodrg]} B.P. Dolan, {\sl Geodesic Renormalisation Group Flow}
hep-th/9511175\hfill\break Int. J. Mod.Phys. {\bf A 12} (1997) 2413 
\smallskip
\item{[\the\Stanley]} H.E. Stanley, Phys. Rev. {\bf 176} (1968) 718
\smallskip
\item{[\the\GuidaMagnoli]} R. Guida and N. Magnoli, Nuc. Phys. {\bf B471} 
(1996) 361
\smallskip
\item{[\the\GrahamHugh]} G.M. Shore, Nuc. Phys. {\bf B362} (1991) 85
\smallskip
\item{[\the\SeibergWitten]} N. Seiberg and E. Witten, Nuc. Phys. {\bf 426}
(1994) 19
\smallskip
\item{[\the\Wallace]} E. Bre\'zin and D.J. Wallace, Phys. Rev. {\bf B7}
(1973) 1967
\smallskip
\item{[\the\ZinnJustin]} J. Zinn-Justin, 
{\sl Quantum Field Theory and Critical Phenomena} \hfill\break
(2nd edition) (1993) O.U.P.
\smallskip
\item{[\the\DenjoeChrisb]} D. O'Connor, C. R. Stephens, A. J. Bray
in {\it Dimensional Crossover in the Large N Limit}
cond-mat/9601146
\item{[\the\Pauli]} W. Pauli, {\sl Theory of Relativity} (1981) Dover
\smallskip
\item{[\the\EllisHawking]} G.F.R. Ellis and S.W. Hawking, {\sl The Large 
Scale Structure of Space-Time}\hfill\break
(1973) C.U.P.
\smallskip
\item{[\the\DenjoeHose]} J. Gaite and D. O'Connor, Phys. Rev. {\bf D54} 
(1996) 5163\hfill\break
J. Gaite, {\sl Relative Entropy, the H-theorem and the Renormalisation Group}
\hfill\break
 hep-th/9610040
\smallskip
\item{[\the\Kullback]} S. Kullback, {\sl Information Theory 
and Statistics}
(1959) Wiley, New York
\smallskip
\item{[24]} B.A. Dubrovin, A.T. Fomenko and S.P Novikov, 
{\sl Modern Geometry - Methods and Applications} Part 1 (2nd edition) (1992) 
Springer
\vfill\eject
\hskip 1truecm
\includegraphics{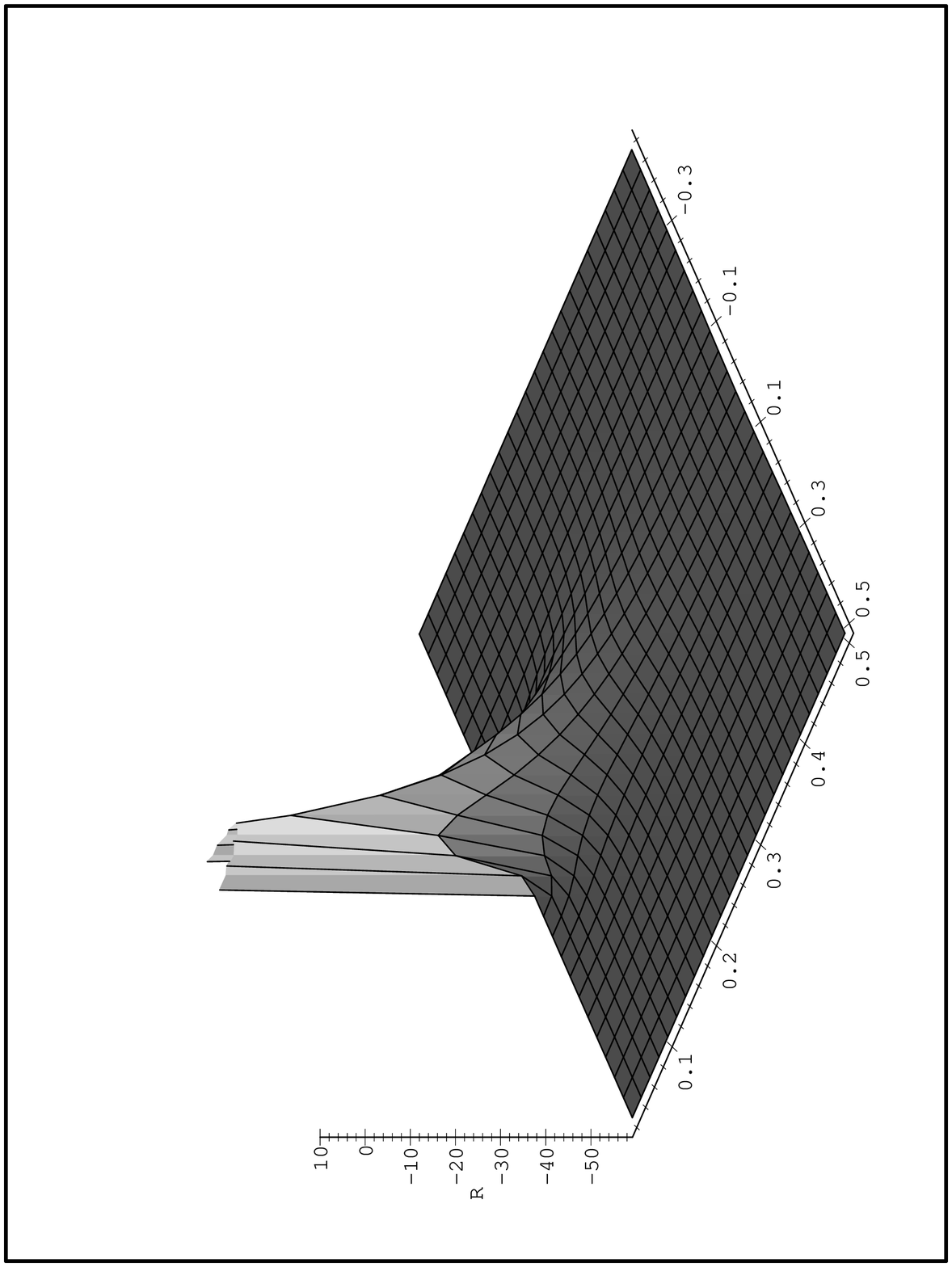}
\centerline{\kern -100pt Fig. 1: Ricci scalar for $\bar\lambda=0.1$}
\raise -180pt\hbox{\hskip 4.5truecm $\bar m$}
\raise -180pt\hbox{\hskip 6.5truecm $\bar\phi$}
\vskip 3cm\hskip 1truecm
\includegraphics{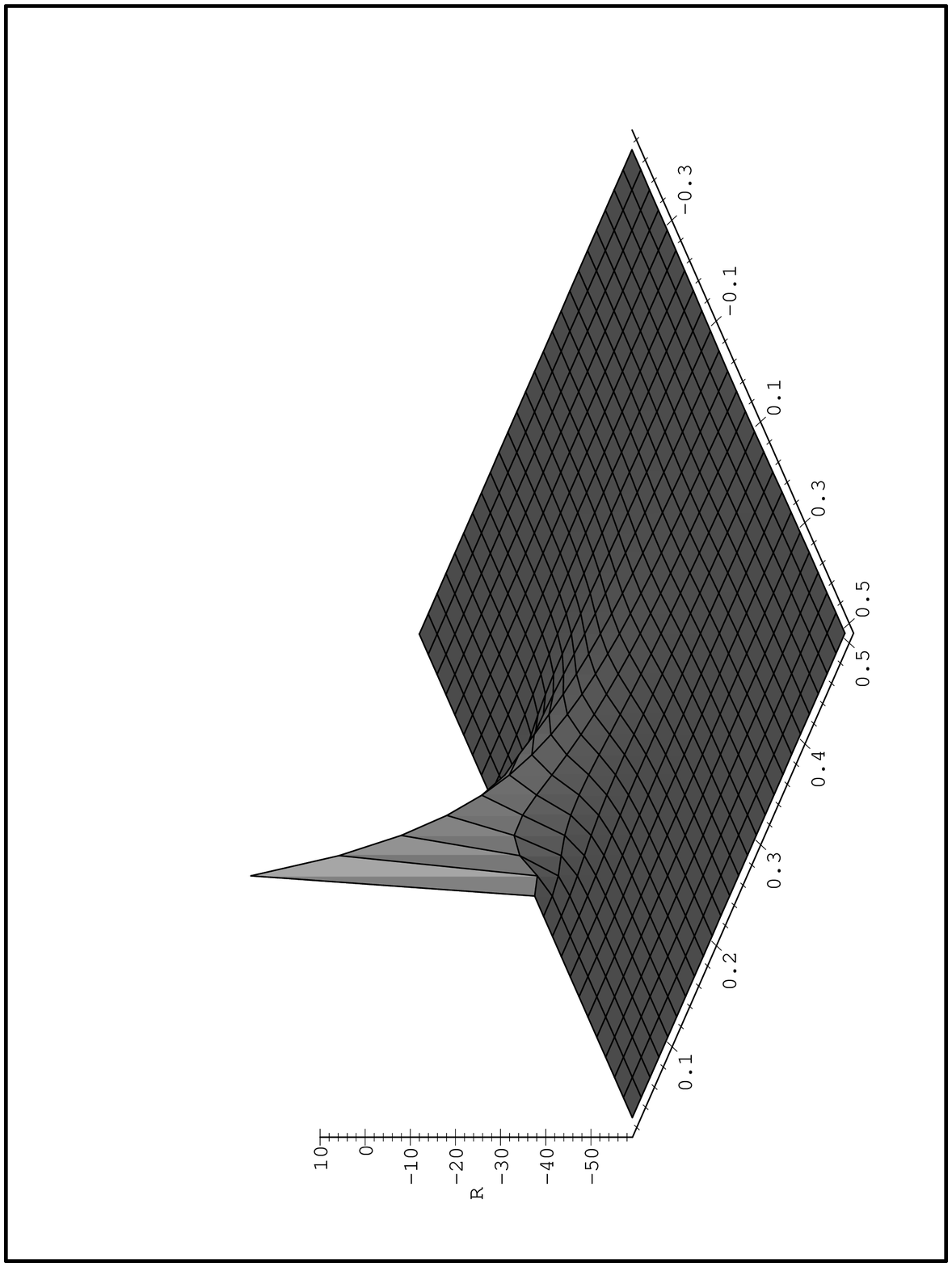}
\centerline{\kern -100pt Fig. 2: Ricci scalar for $\bar\lambda=0.2$}
\raise -180pt\hbox{\hskip 4.5truecm $\bar m$}
\raise -180pt\hbox{\hskip 6.5truecm $\bar\phi$}
\vfill\eject
\hskip 1truecm
\includegraphics{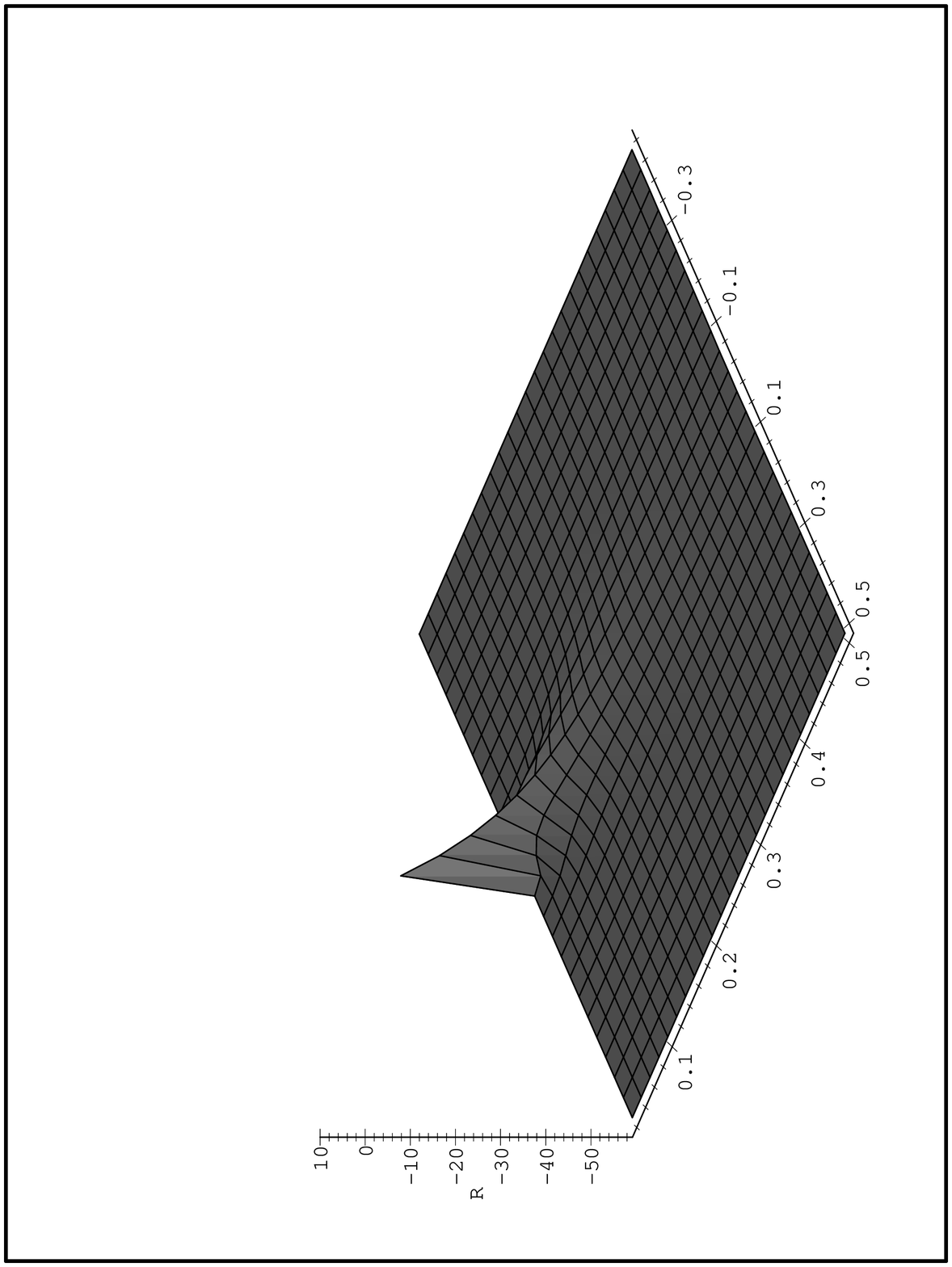}
\centerline{\kern -100pt Fig. 3: Ricci scalar for $\bar\lambda=0.3$}
\raise -180pt\hbox{\hskip 4.5truecm $\bar m$}
\raise -180pt\hbox{\hskip 6.5truecm $\bar\phi$}
\vskip 3cm\hskip 1truecm
\includegraphics{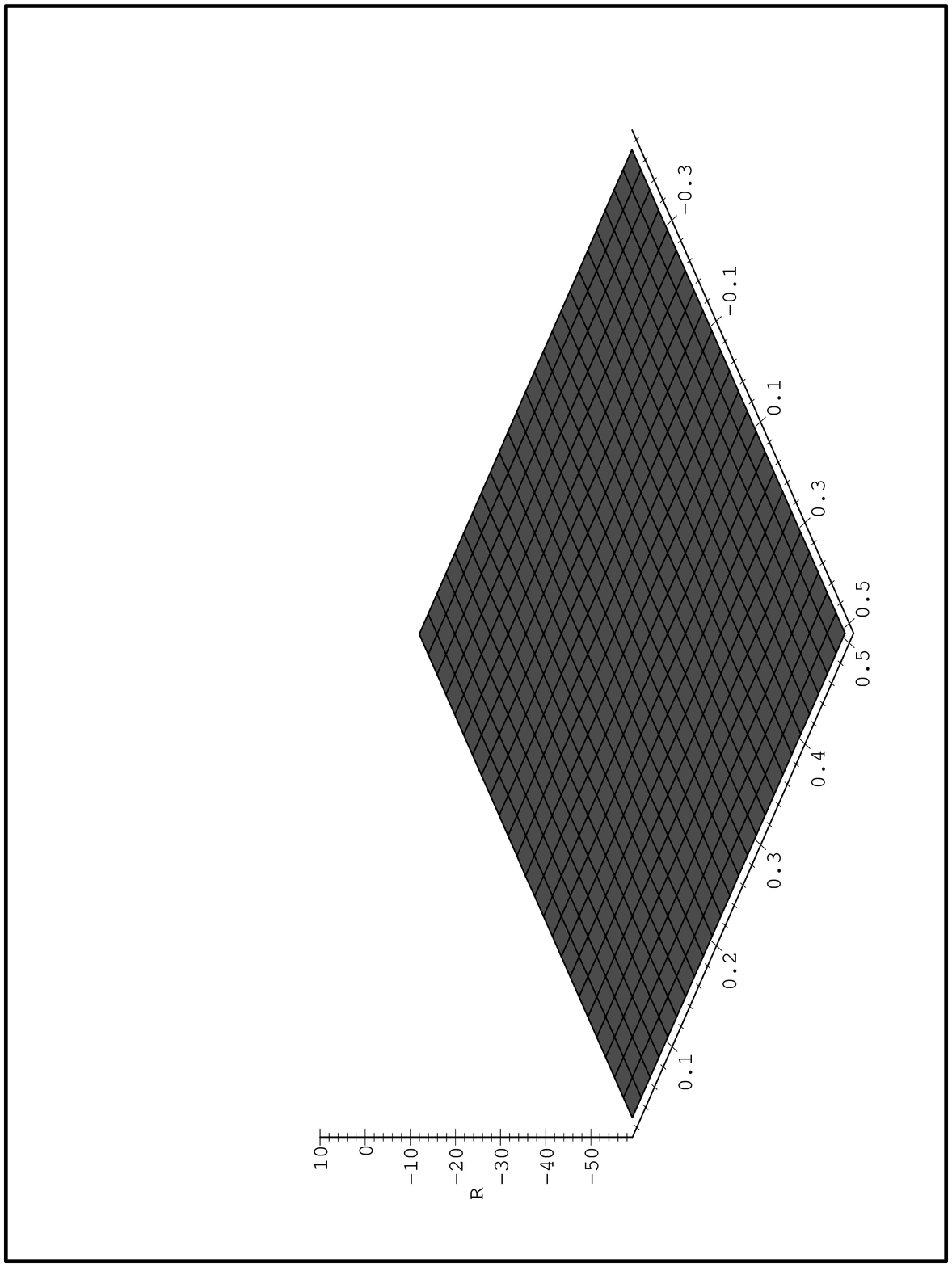}
\centerline{\kern -100pt Fig. 4: Ricci scalar for $\bar\lambda=5.0$}
\raise -180pt\hbox{\hskip 4.5truecm $\bar m$}
\raise -180pt\hbox{\hskip 6.5truecm $\bar\phi$}
\vfill\eject
\hskip 1truecm
\includegraphics{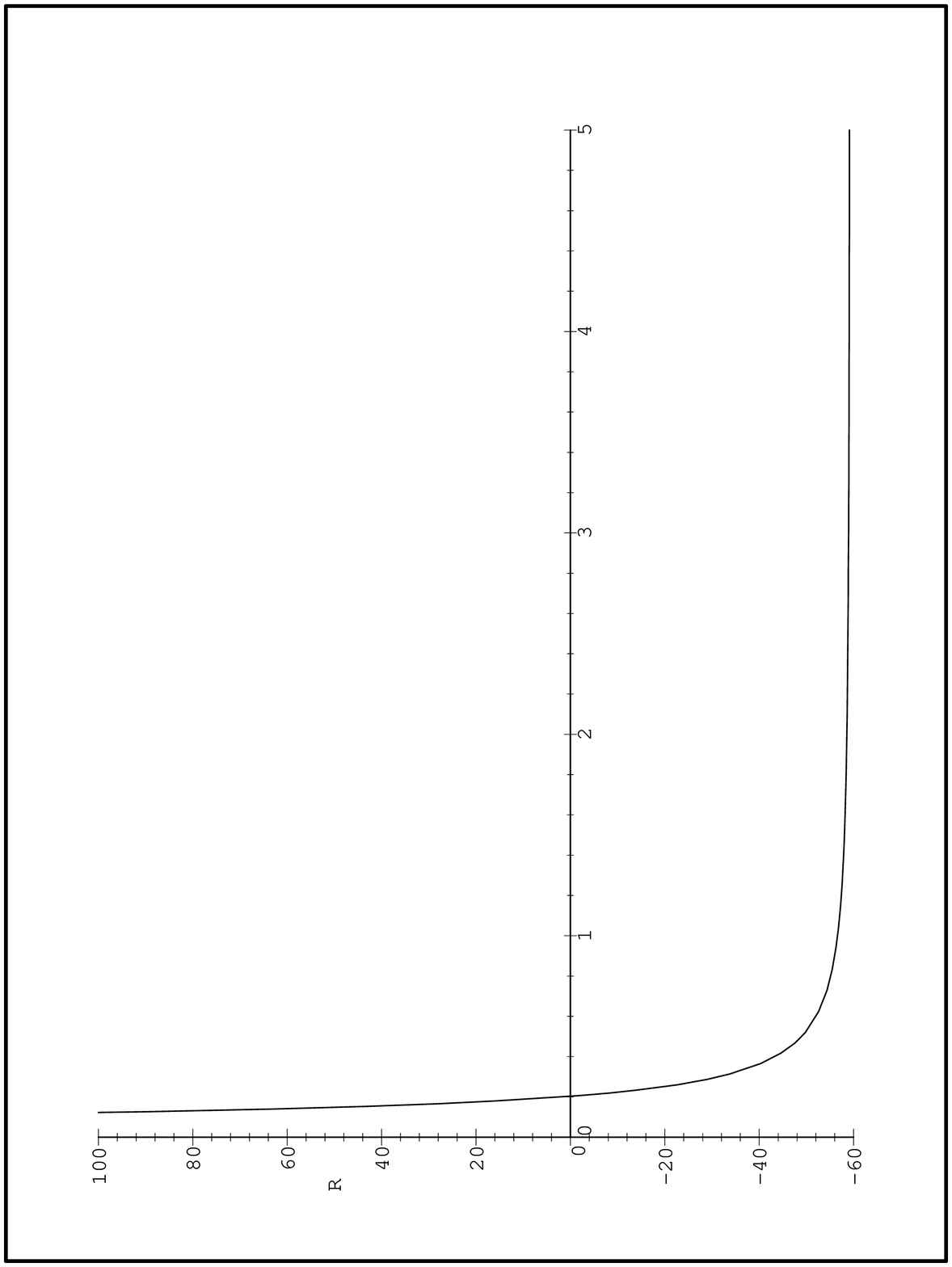}
\raise 10pt\hbox{\centerline{\kern -100pt 
Fig. 5: Ricci scalar along the critical line $\bar\phi=0, \bar m=0$}}
\raise -134pt\hbox{\hskip 8truecm $\bar\lambda$}
\vskip 5cm\hskip 1truecm
\centerline{\kern -100pt Fig. 6: Ricci scalar for $\bar m=0$ and a fixed
value of $\bar\lambda$.}
\centerline{There is a discontinuity at $\bar\phi=0$ reflecting
non-analyticity at the critical line.}
\includegraphics{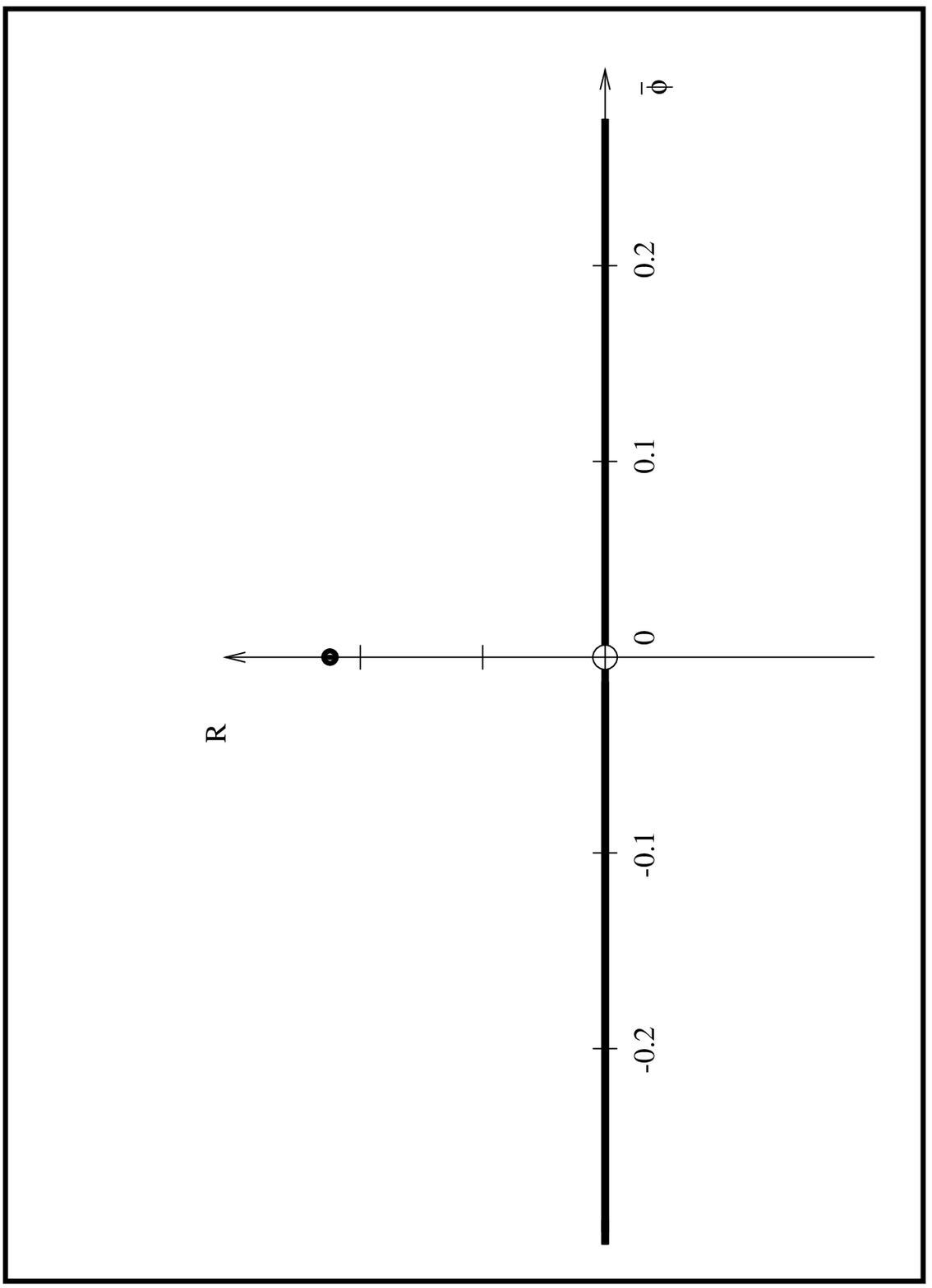}

\vfill\eject
\end